\documentclass[prd,rd,showpacs,twocolumn]{revtex4-1}
\usepackage{amsfonts,amsmath,amssymb,amsthm}
\usepackage[bookmarks, bookmarksopen, bookmarksnumbered]{hyperref}

\usepackage{graphicx}
\usepackage{color}
\usepackage[utf8]{inputenc}
\usepackage[T1]{fontenc}
\usepackage{lmodern}
\usepackage[right]{rotating}
\usepackage{longtable}
\usepackage{array}
\usepackage{multirow}
\usepackage{paralist}
\usepackage{colortbl}

\newcommand{\codename}[1]{\texttt{#1}}

\newcommand{\abs}[1]{\ensuremath{\left| #1 \right|}}
\newcommand{\sq}[1]{\ensuremath{\left[ #1 \right]}}
\newcommand{\rnd}[1]{\ensuremath{\left( #1 \right)}}

\renewcommand{\i}{\mathrm{i}}

\graphicspath{{../pdf/}{../png/}}

\usepackage[normalem]{ulem}

\begin{document}
\title{Spectral analysis of gravitational waves from binary neutron star merger remnants}
\date{\today}

\author{Francesco \surname{Maione}}
\affiliation{Parma University, Parco Area delle Scienze 7/A, I-43124 Parma (PR), Italy}
\affiliation{INFN gruppo collegato di Parma, Parco Area delle Scienze 7/A, I-43124 Parma (PR), Italy}
\author{Roberto \surname{De Pietri}}
\affiliation{Parma University, Parco Area delle Scienze 7/A, I-43124 Parma (PR), Italy}
\affiliation{INFN gruppo collegato di Parma, Parco Area delle Scienze 7/A, I-43124 Parma (PR), Italy}
\author{Alessandra \surname{Feo}}
\affiliation{Parma University, Parco Area delle Scienze 7/A, I-43124 Parma (PR), Italy}
\affiliation{INFN gruppo collegato di Parma, Parco Area delle Scienze 7/A, I-43124 Parma (PR), Italy}
\author{Frank \surname{L\"offler}}
\affiliation{Center for Computation \& Technology, Louisiana State University, Baton Rouge, LA 70803 USA}

\begin{abstract}
In this work we analyze the gravitational wave signal from
hypermassive neutron stars formed after the merger of binary neutron
star systems, focusing on its spectral features. The gravitational
wave signals are extracted from numerical relativity simulations of
models already considered by 
De Pietri et al. [Phys. Rev. D 93, 064047 (2016)], 
Maione et al. [Classical Quantum Gravity 33, 175009 (2016)], 
and Feo et al. [Classical Quantum Gravity 34, 034001 (2017)], 
and allow us to study the effect of the total baryonic mass
of such systems (from $2.4 M_{\odot}$ to $3 M_{\odot}$), 
the mass ratio (up to $q = 0.77$), and
the neutron star equation of state, both in equal and highly unequal
mass binaries. We use the peaks we find in the gravitational spectrum
as an independent test of already published hypotheses of their physical origin
and empirical relations linking them with the characteristics of the
merging neutron stars. 
In particular, we highlight the effects of the mass ratio, which in the
past was often neglected. We also analyze the temporal evolution of
the emission frequencies. Finally, we introduce a modern variant of
Prony’s method to analyze the gravitational wave postmerger emission
as a sum of complex exponentials, trying to overcome some drawbacks of
both Fourier spectra and least-squares fitting. Overall, the spectral
properties of the postmerger signal observed in our simulation are in
agreement with those proposed by other groups. More specifically, we
find that the analysis of Bauswein and Stergioulas 
[Phys. Rev. D 91, 124056 (2015)] is particularly effective for binaries with very low
masses or with a small mass ratio and that the mechanical toy model of
Takami et al. [Phys. Rev. D 91, 064001 (2015)] provides a
comprehensive and accurate description of the early stages of the
postmerger.
\end{abstract}

\LTcapwidth=\columnwidth


\maketitle

\section{Introduction}
\label{sec:intro}

Gravitational waves (GW) from binary neutron star (BNS) mergers are
the next target for Earth-based interferometric detectors, after the
recent first detection of GW from two binary black hole mergers
\cite{Abbott:2016blz,Abbott:2016nmj,Abbott:2017vtc}.
BNSs are a particularly interesting system to study,
since they are also linked to the electromagnetic signal counterpart to the
GWs \cite{Paschalidis:2013jsa,Palenzuela:2013hu,Ponce:2014sza,Metzger:2014yda},
to be the center for r-process 
nucleosynthesis in material that is dynamically ejected \cite{Goriely:2011vg,Rosswog:1998hy}
macronovae~\cite{Metzger:2010sy,Tanvir:2013pia,Berger:2013wna,Bauswein:2015vxa,Sekiguchi:2015dma,Radice:2016dwd,Rosswog:2016dhy}
from the material ejected during and after the merger,
and are related to short gamma ray bursts~\cite{Barthelmy:2005bx,Rezzolla:2011da,Ruiz:2016rai,Paschalidis:2016agf}
(whose central engine mechanism is still disputed).
Even more importantly, BNS mergers can be
thought as a laboratory to study nuclear physics at the extreme
conditions present in neutron star cores~\cite{Baiotti:2016qnr}. The still
unknown equation of state (EOS) of nuclear matter inside the neutron
star core, at densities higher than at nuclear equilibrium, will
leave an imprint on the GW signal emitted both before and after the
merger.

For the coalescent phase, semi-analytic techniques have
been developed that include effects due to the tidal deformability 
effect of matter EOS, in particular within the effective-one-body (EOB)
formalism~\cite{buonanno:1999effective,damour:2010effective,Hinderer:2016eia,Steinhoff:2016rfi},
and tested and validated using numerical relativity (see for example 
\cite{Baiotti:2011am,Hotokezaka:2015xka,Hotokezaka:2016bzh,Dietrich:2017feu}).
On the other hand, for the merger and postmerger phase, numerical relativity is the only
available tool to study the evolution of the remnant and its GW
emission, in the case in which a (hyper)massive neutron star
\cite{Paschalidis:2016vmz}, or a black hole surrounded by an accretion
disk is formed.  If a neutron star remnant is produced, its GW
emission will still be linked to the neutron star EOS. In particular,
many recent works focused on linking the spectral peaks of postmerger
GW emission with some characteristics of the merging neutron stars~\cite{Bauswein:2011tp,Bauswein:2012ya,Stergioulas:2011gd,read:2013matter,Takami:2014tva,bauswein:2014revealing,bauswein:2015unified,Bauswein:2015vxa,bernuzzi:2015modeling,Lehner:2016lxy,Rezzolla:2016nxn},
such as their radius, compactness or tidal deformability. If the progenitor
masses would be known from the inspiral signal, they could be used to constrains the EOS
together with information about the postmerger peaks.

Despite this large body of previous works, which mainly focuses on constructing
empirical relations between GW spectral features and EOS-related
features, there is still an open debate about the \textit{physical}
origin of the postmerger GW signal, especially the subdominant spectral peaks.

In this work, we analyze the postmerger GW signal from numerical
relativity simulations, of which other aspects were already highlighted our previous
works~\cite{DePietri:2015lya,Maione:2016zqz,Feo:2016cbs}. Our goals here
are to get a clearer picture of the postmerger GW emission mechanisms
and their evolution in time, and to act as an independent test for
empirical relations published in the recent literature, which are
often tested only on the same data used to derive them. Our set of
simulations spans several directions in the relevant parameter space,
investigating the effect of the total baryonic mass (from $M_T = 2.4
M_{\odot}$ of model SLy 1.11vs1.11 to $M_T = 3.2 M_{\odot}$
of model SLy 1.44vs1.44), the mass ratio (up to $q =0.77$, 
which corresponds to the largest mass asymmetry observed in a
BNS system in our galaxy \cite{Martinez:2015mya}), and the high
density EOS in both equal and highly unequal mass systems. In
particular, unequal mass systems were less widely investigated 
for the effect of the mass ratio in the postmerger GW analysis
\cite{Bauswein:2012ya,DePietri:2015lya,Sekiguchi:2016bjd,Lehner:2016lxy,Dietrich:2016hky}. In
this work we will highlight the effect of mass-ratio on the GW
spectral features and evaluate the error of applying empirical
relations developed for equal or close-to-equal mass binaries to
highly unequal mass ones. The presence of this effect was already
emphasized in~\cite{Bauswein:2015vxa} for the existence of empirical relations 
between the peak frequency of the GW signal and the radius of the neutron star.

We also adopt, for the first time, a modern version of Prony's method 
\cite{Hua1990,Berti:2007dg,Berti:2007inspiral,Potts2010,Plonka2014}
to analyze GWs from BNS merger remnants. It is a
promising technique, since it is able to overcome some of the
limitations of both Fourier spectrograms and least-squares fitting. In
particular, the Prony analysis allows us to confirm that the
postmerger GW signal is given by a sum of complex exponential modes,
which could not be clearly identified from the Fourier
spectra alone. We also validate those mode frequencies and their time
evolution with an independent data analysis technique.

The paper is organized as follows: in section \ref{sec:setup} we
briefly describe the simulated initial data and the numerical methods
adopted. In section \ref{sec:results} we present the results of our
analysis. In particular, \ref{sec:f2} contains a comparison of some empirical
relations for the dominant spectral peaks with our data, in section
\ref{sec:f1} we discuss different models for explaining the
subdominant peaks in the spectra, before we introduce
the version of the Prony's method we implemented and the results
obtained applying it to our data in section \ref{sec:prony}. Finally, we
conclude our analysis in section \ref{sec:conclusions}. The
work is completed by one appendix \ref{sec:appendix}, containing characteristics
of our initial data, as well as a comparison of our data with existing, universal
formulas for the postmerger peak frequencies.

Throughout this paper we use a spacelike signature $-,+,+,+$, with
Greek indices running from 0 to 3, Latin indices from 1 to 3, and the
standard convention for summation over repeated indices.  The
computations are performed using the standard $3+1$ split into
(usually) spacelike coordinates $(x,y,z)=x^i$ and a time-like
coordinate $t$.  Our coordinate system $(x^\mu)=(t,x^i)=(t,x,y,z)$
(far-from the origin) are, as it can be checked, almost isotropic
coordinates and (far-from the origin) they would have the usual
measure unit of ``time'' and ``space'' and in particular $t$ is close
to be identified as the time measured from an observer at infinity.

All computations have been done in normalized computational units
(hereafter denoted as CU) in which $c=G=M_\odot=1$.  We report all
results in cgs units except for values of the polytropic constant
$K$, whose unit of measurement depends on the value of the
dimensionless polytropic exponent $\Gamma$, so we report $K$ in the
above defined normalized unit CU). We also report masses in terms of the
solar mass $M_\odot$. The reader should note that, as is usual in
most of the work on this subject we describe matter using the
variable $\rho$ (baryon mass density), $\epsilon$ (specific internal
energy) and $p$, instead of, as usually used in Astrophysics,
$\overline{\rho}$ (energy density), $\overline{n}$ (baryon number
density) and $p$.  Their relation is the following: $\overline{\rho}
= e = \rho (1 + \epsilon)$ and $\overline{n}=\rho/m_B$ ($m_B$ is the
baryon mass).

\section{Initial Models and Numerical Methods}
\label{sec:setup}
The models analyzed in this paper were already considered in
our previous works~\cite{DePietri:2015lya,Maione:2016zqz,Feo:2016cbs}
where a detailed discussion of the employed numerical 
methods, their convergence properties as well as of their properties
can be find.
We report here the general simulation setup and parameters and refer 
to those previous articles for simulation setup details. 
In particular, the resolution used in the simulation here presented ($dx=0.25$ CU $=~370$ m) is coarser than 
the one used in other works but that should not efffect the identification of the peaks 
(see \cite{DePietri:2015lya} for a discussion of the convergence properties of the code).

The simulations were performed using the
\codename{Einstein Toolkit}~\cite{Loffler:2011ay}, an open source, modular code
for numerical relativity based on the \codename{Cactus} framework
\cite{Cactuscode:web,Goodale:2002a}. The evolved variables were
discretized on a Cartesian grid with 6 levels of fixed mesh
refinement, each using twice the resolution of its parent
level. The outermost face of the grid was set at $720 M_{\odot}$
(~$1040$ km) from the center. We solved the BSSN-OK formulation of
Einstein's equations
\cite{Nakamura:1987zz,Shibata:1995we,Baumgarte:1998te,Alcubierre:2000xu,Alcubierre:2002kk},
implemented in the \codename{McLachlan} module \cite{McLachlan:web},
and the general relativistic hydrodynamics equations (GRHD) with
\textit{High resolution shock capturing} methods, implemented by the
public \codename{GRHydro} module
\cite{Baiotti:2004wn,Moesta:2013dna}. In particular, we used a
finite-volume algorithm with the HLLE Riemann solver
\cite{Harten:1983on,Einfeldt:1988og} and the WENO reconstruction
method \cite{liu1994weighted,jiang1996efficient}. The combined use of
WENO reconstruction and the BSSN-OK formulation was found
in~\cite{DePietri:2015lya} to be the
best combination within the \codename{Einstein Toolkit} even at low resolution in
\cite{DePietri:2015lya}. For time evolution, we used the
Method of Lines, with fourth-order Runge-Kutta~\cite{Runge:1895aa,Kutta:1901aa}.

Initial data were generated with the \codename{LORENE} code
\cite{Lorene:web,Gourgoulhon:2000nn}, as irrotational
binaries in the conformal thin sandwich approximation. In this work we
analyze a set of simulations with the SLy EOS
\cite{Gourgoulhon:2000nn} (from ref. \cite{DePietri:2015lya}), equal
mass systems with total baryonic mass from $2.4 M_{\odot}$ to $3.2 
M_{\odot}$ and unequal mass systems with the same total baryonic mass
$M_T = 2.8 M_{\odot}$ and mass ratio up to $q = \frac{M_1}{M_2} =
0.77$. We also study simulations with different EOSs, both in equal
mass systems (with total mass $M_T = 2.8 M_{\odot}$, from
ref. \cite{Maione:2016zqz}) and unequal mass binaries, simulating the
merger of the observed system PSR J0453+1559 (see
ref. \cite{Feo:2016cbs}), the BNS system with the largest mass
asymmetry observed so far in our galaxy \cite{Martinez:2015mya}. The
simulations from ref. \cite{DePietri:2015lya} have an initial distance
between the merging stars of $40$ km, while it was set to $44.3$ km for
the simulations from ref. \cite{Maione:2016zqz,Feo:2016cbs}. More
physical initial data characteristics are reported in the Appendix.

The cold part of the EOS is parametrized as a piecewise polytrope with
7 pieces, following the prescription of
ref. \cite{Read:2009constraints}:
\begin{align}
  P_{cold} &= K_i \rho^{\Gamma_i}\\ 
  \epsilon_{cold} &= \epsilon_i +
  \frac{K_i}{\Gamma_i -1} \rho^{\Gamma_i - 1},
\end{align}
where $\epsilon_i$ and $K_i$ are fixed imposing the continuity of the
zero-temperature pressure and the specific energy density ($P_{cold}$
and $\epsilon_{cold}$ respectively), starting from $K_0$, $\epsilon_0
= 0$ and the (zero-temperature) pressure value at the fixed density
$10^{14.7} g/cm^3$. The four lowest-density pieces are common to all
the adopted EOS, and are taken from the SLy EOS \cite{Douchin00}.  The
three high density pieces, instead, differ for the four EOS models we
compared (two nuclear many-body EOSs, SLy \cite{Douchin01} and APR4
\cite{Akmal:1998cf}, and two relativistic mean-filed EOSs, H4
\cite{Lackey:2005tk} and MS1 \cite{Muller:1995ji}). All the
EOS-specific parameters are reported in table \ref{tab:eos}. During
the evolution, the cold EOS is supplemented by an ideal-fluid thermal
component, to ensure thermodynamic consistency in the presence of
shocks. It takes the form of a $\Gamma$-law, with the choice
$\Gamma_{th} = 1.8$ \cite{bauswein:2010testing}.
\begin{equation}
  P_{th} = \Gamma_{th} \rho (\epsilon - \epsilon_{cold}).
\end{equation}

{\renewcommand{\arraystretch}{1.3}
\begin{table}
 \newcolumntype{R}{>{\columncolor[gray]{0.9}}c}
\begin{tabular}{cRcccc}
\multirow{2}{*}{i} & & \multicolumn{4}{c}{$\Gamma_i$} \\\noalign{\vskip-0.3pt}
                   & \multirow{-2}{*}{$\rho_i [\mathrm{g}/\mathrm{cm}^3]$}
                   & APR4 & SLy & H4 & MS1 \\\noalign{\vskip-0.3pt}
\hline
0 & -                      &\multicolumn{4}{c}{1.584}\\\noalign{\vskip-0.3pt}
1 & $2.440 \times 10^{7}$  &\multicolumn{4}{c}{1.287}\\\noalign{\vskip-0.3pt}
2 & $3.784 \times 10^{11}$ &\multicolumn{4}{c}{0.622}\\\noalign{\vskip-0.3pt}
3 & $2.628 \times 10^{12}$ &\multicolumn{4}{c}{1.357}\\\noalign{\vskip-0.3pt}
\multirow{2}{*}{4} &       &2.830&3.005&2.909&3.224\\*[-.4em]
  & $(\rho_4/10^{14})$
  & \multicolumn{1}{>{\columncolor[gray]{0.9}[5pt]}c}{$(1.512)$}
  & \multicolumn{1}{>{\columncolor[gray]{0.9}[5pt]}c}{$(1.462)$}
  & \multicolumn{1}{>{\columncolor[gray]{0.9}[5pt]}c}{$(0.888)$}
  & \multicolumn{1}{>{\columncolor[gray]{0.9}[5pt]}c}{$(0.942)$}\\\noalign{\vskip-0.3pt}
5 & $1\times10^{14.7}$   &3.445&2.988&2.246&3.033\\\noalign{\vskip-0.3pt}
6 & $1\times10^{15}$     &3.348&2.851&2.144&1.325\\\noalign{\vskip-0.3pt}
\end{tabular}
\caption{Parameters for 4 different piecewise polytropic EOSs. $K_0$ for
all EOSs is $6.801 \times 10^{-11}$, with the other $K_i$ chosen to obtain
continuous EOSs. As can be seen, all EOSs show the same low-density behavior,
but start to differ above $\rho_4$ (which is also different for all EOSs).
While $\rho_5$ and $\rho_6$ is the same for all EOSs, they use a quite different
$Gamma_i$ for this high-density regime.}
\label{tab:eos}
\end{table}
}

\subsection{Gravitational waves extraction}
During the simulations the GW signal is extracted
computing the Newman-Penrose scalar $\Psi_4$
\cite{Newman:1961qr,Baker:2001sf} (using the code module
\codename{WeylScalar4}), which is linked to the GW
strain by the following relation, valid only at spatial infinity:
\begin{equation}
  \Psi_4 = \ddot{h}_+ - i \ddot{h}_{\times}, \label{eq:psi4}
\end{equation}
where $h_+$ and $h_{\times}$ are the two polarizations of the complex
GW strain $h = h_+ i h_{\times}$. The signal is then
decomposed in spin-weighted spherical harmonics of weight $(-2)$
\cite{Thorne:1980ru} (by the module \codename{Multipole}):
\begin{equation}
 \psi_4(t,r,\theta,\phi)  = 
 \sum_{l=2}^{\infty} {\sum_{m=-l}^{l} {\psi_4^{lm}(t,r)\ {{}_{-2}\!}{Y}_{lm}(\theta,\phi)}}.
\end{equation}
Since in this work we only focus on the dominant $l = m = 2$ mode,
we will identify $h = h_{2,2}$ for the rest of this paper. In order to
get the GW strain form $\Psi_4$ and minimizing the
extraction errors, one has to extrapolate the signal
extracted within the simulation at finite distance from the source to infinity,
in order for eq. (\ref{eq:psi4}) to be valid. Then, the extrapolated $\Psi_4$
is integrated twice in time, employing an appropriate
technique to reduce the amplitude oscillations caused by
high-frequency noise aliased in the low-frequency signal and amplified
by the double integration process~\cite{Reisswig:2011notes}. We
adopted the procedure developed and extensively discussed in
ref. \cite{Maione:2016zqz}: first, $\Psi_4$ is extrapolated
to spatial infinity using the second order perturbative correction of
Nakano and collaborators~\cite{Nakano:2015perturbative}:
\begin{align}
  r\psi_4^{lm}(t_\mathrm{ret})\left|_{r=\infty}\right. 
  & =\ \left( 1-\frac{2M}{r}\right) \bigg( r\ddot{\bar{h}}(t_\mathrm{ret},r) + \label{eq:extrap}\\
  & - \ \frac{(l-1)(l+2)}{2r} \dot{\bar{h}}(t_\mathrm{ret},r)\            \notag \\
  & + \ \frac{(l-1)(l+2)(l^2+l-4)}{8r^2} \bar{h}(t_\mathrm{ret},r)\bigg). \notag
\end{align}
Both the GW strains at finite radius, which are
present in eq. (\ref{eq:extrap}), and the final extrapolated strain are
computed first by integrating the Newman-Penrose
scalar twice in time with a simple trapezoid rule, starting from zero coordinate
time, and fixing only the two physically meaningful integration
constants $Q_0$ and $Q_1$ by subtracting a linear fit
of itself from the signal:
\begin{align}
 \bar{h}^{(0)}_{lm}\ &=\ \int_{0}^t{dt' \int_{0}^{t'}{dt'' \psi_4^{lm}(t'',r)}} \label{eq:time_int}\\
\bar{h}_{lm}\ &=\ \bar{h}^{(0)}_{lm}\ -\ Q_1 t\ -\ Q_0.\label{eq:fit_int}
\end{align}
Only after the integration, a digital high-pass
Butterworth filter is applied, designed to have a maximum amplitude reduction of
$0.01$~dB at the initial GW frequency $f_{t_0}$ (assumed to be two times
the initial orbital angular velocity, as reported by the LORENE code),
and an amplitude reduction of $80$~dB at frequency $0.1 f_{t_0}$.

All the GW related information will be reported in
function of the retarded time
\begin{align}
  t_\mathrm{ret} &= t - R^*\\
  R^* &= R + 2M_\mathrm{ADM} \log\left(\frac{R}{2M_\mathrm{ADM}} - 1\right).\notag 
\end{align}
From the GW strain obtained with the aforementioned procedure, 
we arrive at the GW amplitude spectral density $\abs{\tilde{h}(f)}f^{1/2}$, 
which is the physical observable we are focusing on in the analysis presented in this work, with:
\begin{equation}
\abs{\tilde{h}(f)} = \sqrt{\frac{\abs{\tilde{h}_+(f)}^2 + \abs{\tilde{h}_{\times}}^2}{2}}, \label{eq:fft}
\end{equation}
where $\tilde{h}(f)$ is the Fourier transform of the complex GW strain:
\begin{equation}
\tilde{h}(f) = \int_{t_i}^{t^f}{h(t)e^{-2\pi \i ft} dt}.
\end{equation}

\section{Results}
\label{sec:results}
\begin{figure*}
  \begin{centering}
    \includegraphics[width=.95\textwidth]{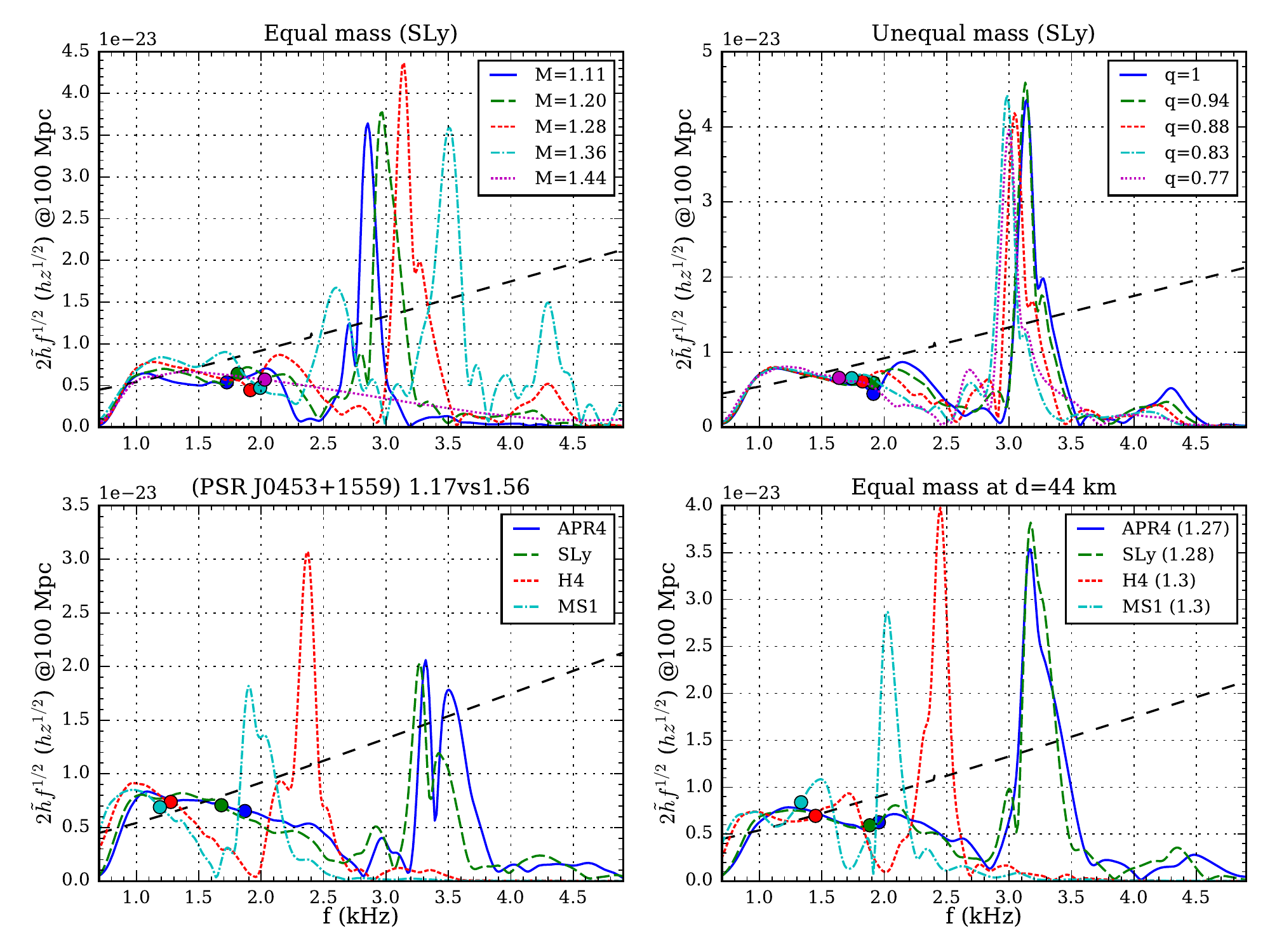}
  \end{centering}
  \caption{Amplitude of the spectral density of the GW signal
    $\abs{\tilde{h}(f)}f^{1/2}$, computed with eq. (\ref{eq:fft})
    for an optimally aligned source at $100$~Mpc. The Fourier
    transform is taken from $8$~ms before to $15$~ms after the merger.
    Filled circles mark the instantaneous frequency at merger.
    The top-left panel shows equal mass models with the SLy
    EOS and different total masses. On the top-left, unequal mass models
    are shown with the same EOS and a fixed total baryonic mass of $M_T = 2.8
    M_{\odot}$. The bottom panels show models with different EOSs,
    reproducing the observed PSR J0453+1559 system (left), or with
    baryonic mass $M = 1.4 M_{\odot}$ for each star (right).
    The filled circles mark the instantaneous frequency at merger.
    The
    dashed black line shows the Advanced Ligo design sensitivity curve
    in the ``zero detuning - high power'' configuration~\cite{TheLIGOScientific:2014jea}.}
    \label{fig:spectrum}
\end{figure*}
\begin{figure*}
  \begin{centering}
    \includegraphics[width=0.95\textwidth]{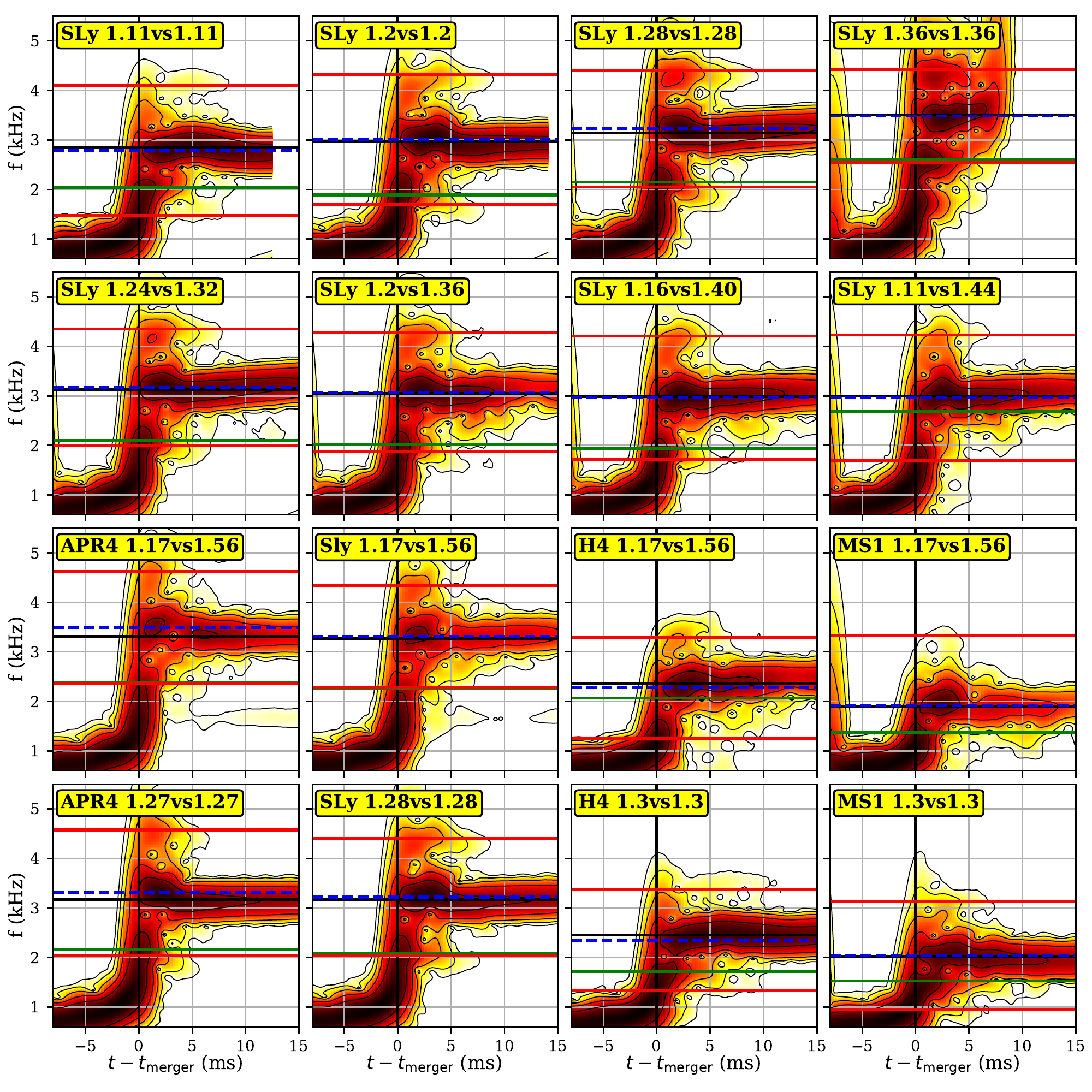}
  \vspace{-7mm}
  \end{centering}
  \caption{Fourier spectrograms of all the not promptly collapsing models represented 
      in figure \ref{fig:spectrum}. The Red lines show the combination  of the 
      dominant peak frequency $f_{2i}$ reported in table~\ref{tab:f2} (dashed-blue) with the 
      oscillation frequency $f_0$ reported in table~\ref{tab:f1}, namely $f_{2i} \pm f_0$
  \label{fig:spectrograms}}
\end{figure*}

In order to study the spectral features of GWs emitted
by the hypermassive neutron star remnant after the merger of BNS
systems, we first compute the whole Fourier spectrum,
as described in the previous section, from $8$~ms before up to
$15$~ms after merger. The results are shown in figure~\ref{fig:spectrum}.
For all models, the spectrum has an initial
growth and a maximum (corresponding to the inspiral GW emission, which
has finite temporal length).
The filled circles in figure~\ref{fig:spectrum}
mark the instantaneous frequency at merger, computed as $f_i =
\frac{1}{2\pi}\left.\frac{d\Phi_\mathrm{GW}(t)}{dt}\right|_{t_\mathrm{merger}}$,
where $t_\mathrm{merger}$ is taken as the
time at which the GW amplitude is maximum, and $\Phi_\mathrm{GW}(t) =
\arctan{\frac{h_{\times}}{h_+}}+2k\pi$ is the accumulated GW phase,
with the integer $k$ chosen to impose its continuity. The segment of
each spectrum at frequencies greater than the merger frequency is
generated by the merger remnant GW emission. In particular, for all
models, it shows a well known dominant peak. This peak corresponds to
the frequency $f_2$ (also called $f_\mathrm{peak}$ or $f_\mathrm{p}$ in the
literature), of the fundamental quadrupolar $m=2$ oscillation mode of
the bar-deformed neutron star formed after the merger
\cite{Stergioulas:2011gd}. Its frequency has been correlated with
different characteristics of the merging neutron stars \cite{Bauswein:2011tp,Bauswein:2012ya}, 
in particular for constructing empirical relations to constrains the neutrons star
EOS with future BNS postmerger GW detections. For a quantitative
discussion of some of those relations, see sec. \ref{sec:f2}.  Most
models also show one or more subdominant peaks, at frequencies both
lower and higher than the dominant one. The scientific debate about
their physical origin is still open (see
ref. \cite{Bauswein:2015vxa,Rezzolla:2016nxn} for an overview of the
most recent results). Like the dominant peak, especially the low-frequency
subdominant peak has been the target for empirical relations
linking it to the characteristics of static stars with the same EOS
as that of the merger remnant. A detailed discussion about subdominant peaks
is presented in section~\ref{sec:f1}. Looking at figure~\ref{fig:spectrum},
one can have a first qualitative impression about
the dependency of the spectral features on the total mass of the
binary, the mass ratio, and the neutron star EOS. In the top-left
panel, equal mass systems with the same EOS (SLy) and different total
mass are compared. With increasing mass, the dominant peak gains more
power and moves towards higher frequencies, as it is expected from a
more compact remnant. For most models, there are two subdominant peaks which are
about equidistant from the dominant peak. They also gain more power with
increasing total mass. As an exception, there is no
recognizable peak with a frequency higher than the dominant one for the
system SLy 1.11bs1.11 with $M_T = 2.4 M_{\odot}$. The highest-mass system considered
($M_T = 3 .2M_{\odot}$, model SLy 1.36vs1.36, which collapses to black-hole $7$~ms after the
merger) instead, shows an additional low frequency peak, which does
not correspond to any of the emission mechanisms analyzed so far in
the literature. It is situated at a lower frequency than the merger one, but
it comes, nevertheless, from the postmerger, as confirmed by the
spectrogram (see later in the text and figure
\ref{fig:spectrograms}). The top-right panel shows unequal mass
systems with the same total baryonic mass ($M_T = 2.8 M_{\odot}$) and
EOS (SLy), but different mass ratios, up to $q = 0.77$. As already
reported in ref.~\cite{DePietri:2015lya} and confirmed in other works
published in the past year~\cite{Lehner:2016lxy,Dietrich:2016hky},
the mass asymmetry leads to a lower dominant peak
frequency and subdominant peaks with progressively less power. The
bottom panels show the EOS effects for unequal (left) and equal (right)
mass binaries: the softest EOSs (SLy and APR4), which lead to the most
compact remnants, are characterized by a dominant peak at higher
frequencies (therefore, more difficult to detect in current generation
GW interferometers). The
low-frequency subdominant peaks have relatively higher power
for the less compact stars (H4 and, in a more pronounced fashion,
MS1), while a high frequency subdominant peak is clearly recognizable
only in the most compact stars. This difference was already noted in
the unified model of~\cite{bauswein:2015unified}, and was being attributed
to the possibility of different emission mechanisms being responsible for
the subdominant spectral peaks in soft and stiff EOS stars (see
section~\ref{sec:f1} for a deeper discussion about this hypothesis).

References.~\cite{Clark:2015zxa,Rezzolla:2016nxn,Dietrich:2016hky} highlighted
the importance of not only analyzing the spectrum taken over the entire merger time, but also 
analyzing its time evolution. This is important, because the signal components are not
fixed in time, but their frequencies evolve dynamically. Figure~\ref{fig:spectrograms}
shows the GW Fourier spectrogram for all the
analyzed models, computed taking the Fourier transform in intervals of
$5$~ms, with a superposition of 95\%. Before the Fourier transform,
the time domain signal is first padded with zeros to obtain a better
frequency resolution, with a padding length of twice the length of the original
signal. The first qualitative information noticeable
in the spectrograms is that the subdominant modes are
short lived, and that they decay during the first $5$~ms after
merger. Even if they are too weak to be fully visible in the
spectrograms, all the subdominant modes are active just after the merger. 
For unequal mass systems with a soft EOS this show that the main emission 
mechanism is only suppressed by the mass asymmetry, but it is 
nevertheless active. Finally, the dominant emission peak does not show a 
fixed frequency, but changes slightly with time. 

In the following subsections we present the results obtained 
for the descriptions of the postmerger spectrum of binary neutron star mergers.

\subsection{Rapid change of $f_2$ within few ms after merger}
  The frequency of the main emission mode rapidly changes in the
  first milliseconds after the merger, when also the subdominant modes
  are active, and when the merger remnant is rapidly evolving toward a
  more stable, equilibrium configuration. 
  This feature was already noted in ref.~\cite{Rezzolla:2016nxn}
  and was described using the notation $f_{2i}$ to 
  indicate a short-lived mode that evolves into the $f_2$ frequency. 
  This continuous process seems to have different
  characteristics depending on the binary physical characteristics as was observed 
  in the spectrogram of~\cite{Clark:2015zxa}. For
  example, in some models this frequency change is a slow process,
  with the dominant frequency first increasing (after the merger) and
  then decreasing, to reach its quasistationary stage value, as
  happens for the equal mass $M=1.11 M_{\odot}$ SLy EOS model (top-left panel),
  or for the equal mass model with
  the H4 EOS (third panel of the bottom row). In other systems this
  maximum in the dominant frequency is also present, but the
  quasi-stationary phase is reached much more quickly, as the equal
  mass systems with the SLy EOS and $M = 1.20, 1.28 M_{\odot}$ (second
  and third panels in the first row) or all the systems
  with the parameter of the observed PSR J0453+1559 binary (third
  row). Finally, there are systems for which the dominant frequency
  just decreases from its value right after the merger to the
  quasistationary phase value (like the equal mass APR4 model).  It
  is interesting to note that in the PSR J0453+1559 system with a soft
  EOS (SLy or APR4), the change in frequency of the dominant peak is
  visible also in the full spectrum (figure \ref{fig:spectrum},
  bottom-left panel), as the splitting of the main peak, and as already
  reported in~\cite{Feo:2016cbs}. Although this could be considered an artifact of
  the short post merger simulated time, it is unlikely to change
  even for a longer observation period. The reason is that it is
  caused by the fact that the GW amplitude quickly decreases after the
  first transient phase in high mass asymmetry models, allowing for
  the short but with relative high amplitude transient emission to
  leave an imprint on the overall spectrum as a local maximum, which
  does not happen in equal or close-to-equal mass binaries, where the
  GW emission amplitude decreases more slowly in the quasi-stationary
  phase. 
\subsection{Slow increase of $f_2$ at late times}
The dominant mode frequency, in the last part of the signal,
  when it is the only active component, increases with time in
  most models. This effect is expected, because the angular momentum
  emitted in GW and redistributed by hydrodynamical
  processes drives the merger remnant to be more axisymmetric (damping
  the amplitude of the GW emission) and more compact (increasing its
  frequency). This frequency increase is more pronounced in models
  with higher total mass (see first row of
  fig. \ref{fig:spectrograms}) and in equal mass models, while highly
  unequal mass ones show little or no frequency change in the temporal
  interval considered in this work, independently from the EOS (see
  third row). This is quite easy to understand, since the remnants of
  unequal mass binary mergers are less compact, and, in particular,
  more matter is ejected far from its core due to the tidal
  deformation of the lower mass star by its companion gravitational
  field which begins in the last orbits before the merger. 

\subsection{Other features}
Finally, other subdominant modes are visible from the spectrograms,
which are not explained by any of the standard pictures published so
far, like the already mentioned low-frequency peak in the collapsing
model SLy1.36vs1.36, which is
generated at merger and lasts for the first $4$~ms, or the extended
low frequency emissions around $1.7$~kHz visible in the PSR J0453+1559
system with the APR4 or EOSs, which start to develop about $10$~ms
after the merger, already noticed and discussed in
ref.~\cite{Feo:2016cbs}.

\subsection{The dominant emission frequency $f_2$ and its link with the stellar properties}
\label{sec:f2}

\begin{table}
  \begin{tabular}{rl@{\hskip 1.5em}ccccc}
    \multicolumn{2}{c}{\multirow{2}{*}{Model}} & $f_{2i}$ & $f_2$ & $f_{2}^B$ \cite{Bauswein:2015vxa}    & $\Delta R_{M=1.6}$ & $f_2$ \cite{Lehner:2016lxy} \\
                                              && [kHz]    & [kHz] & [kHz]                &  [km]              &  [kHz]  \\   
    \hline
    SLy &1.11vs1.11  & 2.79 & 2.85 & 2.784 & 0.16 & 2.83\\ 
    SLy &1.20vs1.20  & 3.01 & 2.96 & 3.009 & 0.10 & 3.03\\
    SLy &1.28vs1.28  & 3.23 & 3.14 & 3.212 & 0.15 & 3.20\\
    SLy &1.36vs1.36  & 3.48 & 3.51 & 3.410 & 0.19 & 3.38\\
    \hline
    SLy &1.24vs1.32  & 3.18 & 3.13 & 3.210 & 0.17 & 3.20\\
    SLy &1.20vs1.36  & 3.07 & 3.05 & 3.210 & 0.35 & 3.20\\
    SLy &1.16vs1.40  & 2.97 & 2.98 & 3.210 & 0.49 & 3.20\\
    SLy &1.11vs1.44  & 2.97 & 3.00 & 3.197 & 0.43 & 3.20\\
    \hline
    APR4 &1.17vs1.56 & 3.49 & 3.32 & 3.574 & 0.50 & 3.67\\
    SLy  &1.17vs1.56 & 3.31 & 3.27 & 3.427 & 0.31 & 3.41\\
    H4   &1.17vs1.56 & 2.27 & 2.37 & 2.503 & 0.44 & 2.25\\
    MS1  &1.17vs1.56 & 1.91 & 1.90 & 2.179 & 2.30 & 1.88\\
    \hline
    APR4 &1.27vs1.27 & 3.31 & 3.17 & 3.336 & 0.35 & 3.47\\
    SLy  &1.28vs1.28 & 3.22 & 3.17 & 3.212 & 0.08 & 3.20\\
    H4   &1.30vs1.30 & 2.35 & 2.45 & 2.382 & 0.21 & 2.12\\
    MS1  &1.30vs1.30 & 2.03 & 2.02 & 2.081 & 0.29 & 1.80\\
  \end{tabular}
  \caption{Dominant peak frequency, measured from the full spectrum
    $f_2$, or from the spectrum up to $5$~ms after the merger
    $f_{2i}$, taking the maximum of the corresponding amplitude
    spectral density after interpolating it with a cubic spline with
    resolution $1$~Hz. The values here are slightly
    different from the ones in~\cite{DePietri:2015lya} due to the
    different methodology for computing $f_2$ from the data (in the
    cited paper it was computed using a fit of the time domain signal),
    and the different time interval used. In addition, the results of
    simulations with SLy EOS and $M=1.4 M_{\odot}$ show
    different values due to
    different initial stars distances (see~\cite{Maione:2016zqz} for a
    detailed study about its influence) and the different symmetries
    imposed during the evolution. However, they are still fully compatible
    within the discrete Fourier transform error ($47$ Hz). The
    fourth column reports the predicted value for $f_2^B$ using the
    empirical relation of~\cite{Bauswein:2015vxa}. The fifth column
    reports the error in the determination of the radius of a $M=1.6
    M_{\odot}$ static neutron star using the aforementioned relation
    and the real $f_2$ value measured from our data. Finally, the last
    column shows the predicted peak frequency with the relation of
    \cite{Lehner:2016lxy}.}
  \label{tab:f2}
\end{table}
  
The physical mechanism behind to the dominant peak in the
postmerger GW spectrum is well known and agreed upon in the
literature. As anticipated in the Introduction, several empirical
relations have been developed to link the peak frequency $f_2$ with
the merging star characteristics, like their radii
\cite{bauswein:2012measuring,hotokezaka:2013remnant,bauswein:2014revealing,Bauswein:2015vxa},
compactness
\cite{Takami:2015gxa,Takami:2014tva,Lehner:2016lxy}, or tidal deformability
\cite{read:2013matter,bernuzzi:2015modeling,Rezzolla:2016nxn,Bose:2017jvk}. One of
the purposes of this work is to use our data, which cover a relevant
portion of the expected BNS parameter space, as an independent test
for such relations, in order to check their validity and estimation
error on a set of simulations different from the ones used to obtain
the relation parameters with nonlinear fitting.

In particular, we
started form the results of ref.~\cite{Bauswein2012a}, which state that
for equal mass models, $f_2^B$ correlates tightly with the radius of a static neutron star in
equilibrium with the same EOS and a mass higher than the mass of each
merging star. In particular, the dominant postmerger emission
frequency from a system with two $1.35 M_{\odot}$ stars was correlated
with the radius of a static star of $M = 1.6 M_{\odot}$. In
ref.~\cite{Bauswein:2015vxa} [Eq.~(2)], a relationship which connects the radius
of a static TOV star with the $f_2$ frequency and the total
gravitational mass [see Eq.~(\ref{formula:f2B}) in Appendix A for $f_2^B$] 
of the merging system was also presented, but was reported to have higher errors in the obtained radii
with respect to the fixed-total-mass relations. However, it is important to keep in
mind that the authors in~\cite{Bauswein:2015vxa} already noted that for 
an unequal mass $q=0.8$ merger, such a relation it is not naturally fulfilled.
We compared the
aforementioned $R(f_2^B,M_g)$ relation with our data of systems with varying total gravitational mass.
The result of such a comparison are
reported in table~\ref{tab:f2}, together with the corresponding errors in the obtained radii.
The  radii of TOV stars of mass $M = 1.6 M_{\odot}$ are computed for each EOS
using the \codename{rns} code~\cite{Stergioulas95}.

We want to stress that eq. (2) of \cite{Bauswein:2015vxa}, like most
empirical relationships of this kind so far, does not take mass
ratio effects into account. According to our results, this can cause an error in
the inferred radius of the order of $500$~m for the largest mass asymmetries
observed in double neutron star systems. To take
the mass ratio into account, a new relation was
developed in~\cite{Lehner:2016lxy}, correlating $f_2$ linearly with the stars contact
frequency, which is sensible to the mass ratio and can be obtained, to
first approximation, from the stars' masses and compactness~\cite{Damour2012}.
Results shown in table~\ref{tab:f2}
show, however, that the dependency between contact frequency and
mass ratio seems to be too weak to fully account for the differences
observed in $f_2$, as was already observed in figure~4 in ref.~\cite{Lehner:2016lxy}.
In essence, this simpler and physically motivated
empirical relation, seems to perform worse on our data
than the correlation with the radius of a $M = 1.6 M_{\odot}$ static star.

\subsection{Physical interpretation and correlations of the subdominant frequencies}
\label{sec:f1}
\begin{figure*}
  \begin{centering}
    \includegraphics[width=0.95\textwidth]{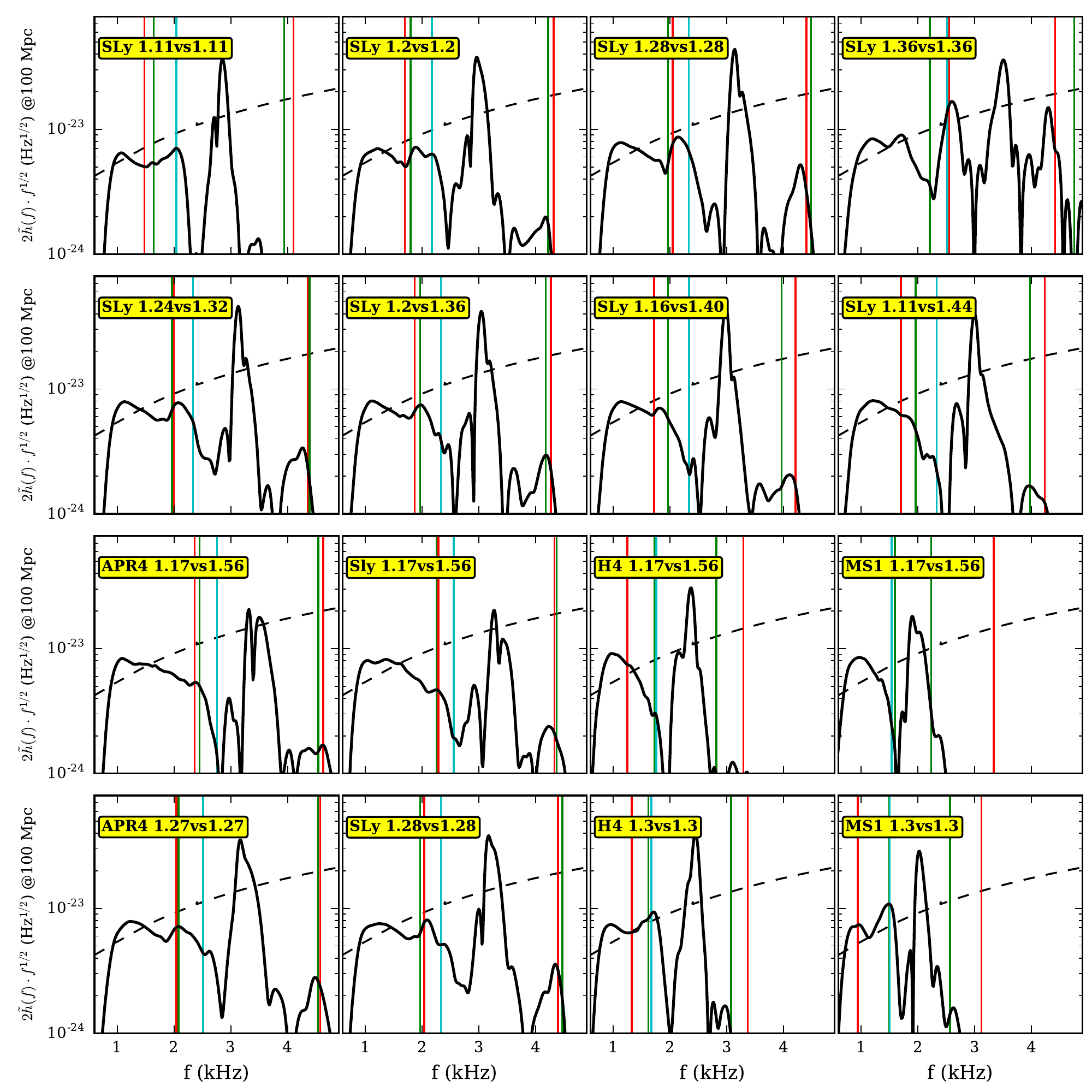}
  \end{centering}
  \vspace{-2mm}
  \caption{
    Amplitude of the spectral densities for all models also presented
    in figure \ref{fig:spectrum}, except the promptly collapsing
    one. 
    The red vertical lines correspond to $f_{2i} - f_0$ and $f_{2i} + f_0$. 
    The vertical green solid lines correspond to the empirical relationship 
    $f_1^T$ of  Eq.~(\ref{eq:f1T}) and the associated frequency 
    $f_3^T=2 f_{2i}-f_1^T$.
    The cyan line shows the application of empirical relationship $f_{spiral}^T$ of Eq.~(\ref{eq:fspiralT}).
    It should be noted that although this relationship does not 
    always match the $f_1$ peak, it seems to correspond to 
    others (even more) subdominant peaks in the spectrum.
    }  \label{fig:AllSpectra}
\end{figure*}

\begin{table}
  \begin{tabular}{rl@{\hskip 1.5em}cccc}
    \multicolumn{2}{c}{\multirow{2}{*}{Model}} & $f_1$ & $f_3$ & $f_0$ & $f_\mathrm{merger}$\\
                                              && [kHz] & [kHz] & [kHz] & [kHz]\\
    \hline
    SLy &1.11vs1.11  & 2.04 & -    & 1.31 & 1.72\\
    SLy &1.20vs1.20  & 1.89 & 4.17 & 1.31 & 1.81\\
    SLy &1.28vs1.28  & 2.15 & 4.30 & 1.19 & 1.91\\
    SLy &1.36vs1.36  & 2.60 & 4.30 & 0.93 & 1.98\\
    \hline          
    SLy &1.24vs1.32  & 2.10 & 4.26 & 1.18 & 1.90\\
    SLy &1.20vs1.36  & 2.01 & 4.19 & 1.20 & 1.82\\
    SLy &1.16vs1.40  & 1.93 & 4.11 & 1.24 & 1.73\\
    SLy &1.11vs1.44  & -    & -    & 1.27 & 1.63\\
    \hline          
    APR4 &1.17vs1.56 & 2.38 & 4.62 & 1.13 & 1.87\\ 
    SLy  &1.17vs1.56 & 2.25 & 4.23 & 1.02 & 1.67\\
    H4   &1.17vs1.56 & -    & -    & 1.02 & 1.27\\
    MS1  &1.17vs1.56 & -    & -    & 1.42 & 1.18\\
    \hline          
    APR4 &1.27vs1.27 & 2.16 & 4.48 & 1.27 & 1.96\\ 
    SLy  &1.28vs1.28 & 2.12 & 4.35 & 1.18 & 1.87\\ 
    H4   &1.30vs1.30 & 1.71 & -    & 1.02 & 1.45\\
    MS1  &1.30vs1.30 & 1.52 & -    & 1.09 & 1.32\\
  \end{tabular}
  \caption{For each model, the second and thirds columns show the subdominant peak
    frequencies $f_1$ and $f_3$, measured taking the local maxima of the amplitude
    spectral density, after interpolating it using a cubic spline with
    a resolution of $1$~Hz. See caption of table~\ref{tab:f2} for an
    explanation about the differences to the numerical values
    reported in our previous
    works~\cite{DePietri:2015lya,Maione:2016zqz,Feo:2016cbs}, and between the
    two simulations with the SLy EOS and $M = 1.4 M_{\odot}$ for each
    star. In the fourth column we report the quasi-radial oscillation
    frequency $f_0$, evaluated taking the peak of the maximum density
    oscillations spectrum, computed in a $10$~ms interval starting at merger,
    resulting in an sensitivity of $100$~Hz. The last column shows the instantaneous
    frequency $f_\mathrm{merger}$ at merger time by taking the derivative of the accumulated
    GW phase at the time of maximal GW strain amplitude.}
  \label{tab:f1}
\end{table}

In the literature have been proposed various explanations for
the subdominant peaks $f_1$ and $f_3$ (also
called $f_-$ and $f_+$) which appear in the spectrum of postmerger GW
emission in most BNS models. 

The first hypothesis but forward  was to consider them as the result 
of the  combinations of the $m=0$ quasi radial oscillation mode and the
fundamental $m=2$ mode \cite{Stergioulas:2011gd} . In most models, the
subdominant peaks are almost equidistant from the dominant one. The
red horizontal lines in figure \ref{fig:spectrograms}, and the
corresponding vertical lines in figure \ref{fig:AllSpectra}, showing
on each panel the GW spectrum of a single model, are drawn at frequencies
$f_{2i} - f_0$, which are the theoretical frequencies of the mode
combination. Here $f_{2i}$, adopting the notation of
\cite{Rezzolla:2016nxn}, is the dominant frequency in the first
milliseconds after the merger, evaluated taking the maximum of the
amplitude spectral density computed up to $5$ ms after the
merger. $f_0$, instead, is the frequency of the quasi-radial
oscillations, computed from the spectrum of the maximum density (or
minimum lapse) oscillations (see figure \ref{fig:rho_spectrum}).  In
most models the frequency predicted for the mode combination is a very
good approximation for the subdominant peaks in the spectrum. However,
it is significantly different in the less compact stars, either low
mass models with a soft EOS (such as the equal mass model with the SLy
EOS and $M = 1.11 M_{\odot}$ for each star, top-left panel in figure
\ref{fig:AllSpectra}), or models with a stiff EOS (such as the stars with
$M \simeq 1.28 M_{\odot}$ and the H4 or MS1 EOS).

Before addressing the mode combination interpretation in
the less compact stars one should consider that 
in \cite{Takami:2015gxa,Takami:2014tva,kastaun:2015properties}
was hypothesized and analyzed the possibility that all the subdominant peaks 
are generated by the modulation of the dominant mode due to the radial 
oscillation of the rotating double core structure formed right after 
the merger and that this modulation could be described 
by a mechanical toy model \cite{Takami:2014tva}.   
According to this interpretation, it is possible to find a single 
relationship connecting $f_1$ to the merging
stars characteristics, and, in particular, to their EOS, since this
subdominant peak is produced by the same mechanism in all models.
A similar relation, fitting $f_1$ with a third order polynomial in the
initial stars average compactness, was developed in
\cite{Takami:2014tva} and refined in \cite{Rezzolla:2016nxn}. Its
predictions, for our data, are reported in figure
\ref{fig:AllSpectra} as the solid green lines.
In almost every model it is able to reproduce well the subdominant peaks,
also for the stiff EOSs, where the mode combination hypothesis failed.
It performs slightly worse than the mode combination hypothesis in the
model close to the collapse threshold (SLy EOS and $M=1.36 M_{\odot}$ 
for each star). In this case, the only models that are not effectively described by 
the proposed universal mechanics are: the lowest equal-mass model with
mass $M = 1.11 M_{\odot}$ (that is quite unlikely to be present in nature)
and some of the extremely unequal mass models, namely, SLy 1.11vs1.44, H4 1.17vs1.56,
and MS1 1.17vs1.56. 

A different possibility was considered in ref. \cite{bauswein:2015unified}
to construct a unified picture. In this case the low frequency GW
subdominant peak in the less compact models (at frequencies
denominated $f_\mathrm{spiral}$ was interpreted as due to the emission from
the spiral arms structure formed after the merger, which rotates
slower than the central double core structure, with a rotation
frequency of $\frac{f_\mathrm{spiral}}{2}$, while the subdominant peaks in
the more compact stars are considered to be produced by the $m=2$ and
$m=0$ mode combination, which, as explained before, is consistent also
with our data.  Indeed, here the word unified should be interpreted 
as the assertion that the two associated peaks are always present 
and that the dominant  $f_1$ peak is just the strongest of the two.
From the data of ref. \cite{bauswein:2015unified},
where the $f_\mathrm{spiral}$ peak was identified in the postmerger emission
of several binary systems with different EOSs, in
\cite{Rezzolla:2016nxn} an empirical relationship was derived,
connecting $f_{spiral}^T$ (see Eq. (\ref{eq:fspiralT})) to the average mass and compactness of the
merging stars, with a second order expression. Its predictions,
applied to our simulations, are shown in figures-\ref{fig:AllSpectra}
by the dash-dotted cyan lines. They agree very well with the low frequency subdominant peak in
the less compact models, where the mode combination cannot explain the
right $f_1$ frequency. In particular, the spectrograms of some models
with intermediate compactness (the equal mass ones with the SLy EOS
and $M = 1.3, 1.4 M_{\odot}$ for each star) show the presence of two
low-frequency subdominant GW emissions, one close to the predicted
frequency of $f_{2i}-f_0$ or $f_1$ from
\cite{Takami:2014tva,Rezzolla:2016nxn}, and the other, at higher
frequency and with a shorter duration, close to the value predicted
for $f^T_{spiral}$. This is consistent with similar results found in
ref. \cite{bauswein:2015unified,Bauswein:2015vxa} for the class of
models they defined \textit{Type II}.

\begin{figure}
  \begin{centering}
    \includegraphics[width=0.48\textwidth]{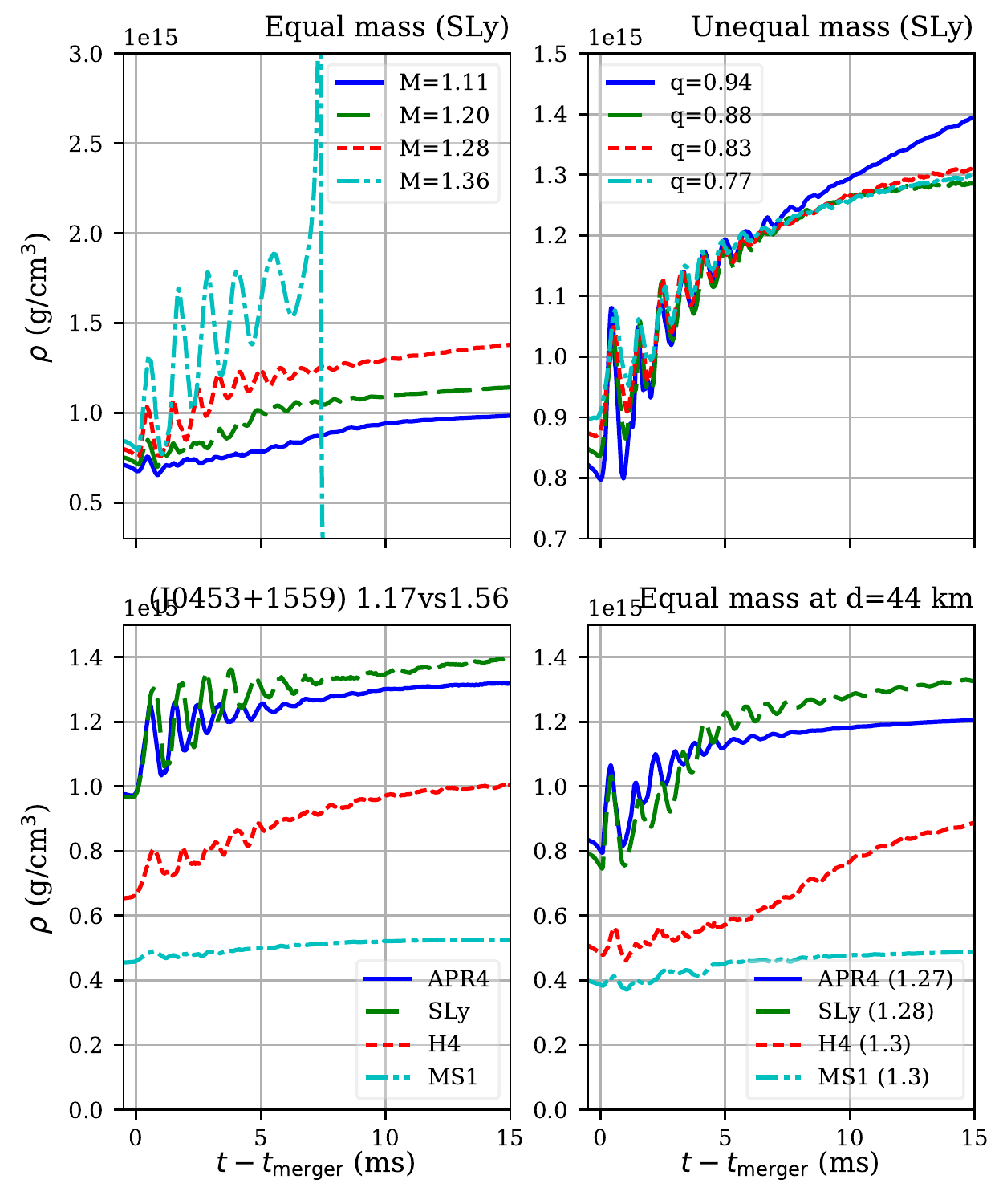}
  \end{centering}
  \vspace{-7mm}
  \caption{Evolution of the maximum density in the merger remnant.}
  \label{fig:rho}
\end{figure}
In order to investigate further these hypothesis, we studied the
evolution of the maximum density and minimum lapse in the
remnant. Figure \ref{fig:rho} shows the maximum density evolution for
all the discussed models. The density grows after the merger, with, in
most models, superimposed oscillations in the first milliseconds. At
the end of the longest simulations, the density reaches a stable
maximum value, when the star is in the quasi-stationary phase. The
maximum density is obviously higher in the more compact stars. In
particular, the dominant effect is due to the EOS. In the stiffest
EOS, the last polytropic piece of our parametrization (at densities
higher than $10^{15} g/cm^3$) is never reached in our models. The
density also appears to grow slower and aim asymptotically to a lower
equilibrium value in systems with a large mass asymmetry. This is easy
to explain, since the remnant of unequal-mass BNS mergers is less
compact due to the tidal deformation of the lower mass star in the
late inspiral and merger phase. Density oscillations, also, have a
higher amplitude in the more compact models, in particular in those
closer to the threshold for quasi-radial collapse (SLy EOS and $M=1.36
M_{\odot}$ for each star), while they seem to have a similar frequency
in almost all models, excluding the aforementioned SLy1.5vs1.5
collapsing model, for which it becomes lower getting closer to the
collapse time. Density oscillations have a negligible amplitude,
instead, in the less compact stars (the ones with MS1 EOS or with SLy
and $M=1.11 M_{\odot}$ for each star), which correspond, also, to the
models for which the subdominant peaks in the spectrum are not well
explained by $m=2$ and $m=0$ mode combination.  In unequal-mass
models, density oscillations are still present, but have a lower
amplitude increasing the mass asymmetry, in particular in the first
2 ms after the merger. This is consistent with their
connection to the subdominant peaks in the GW spectrum: there is still
an emission at frequencies around $f_2$ (as seen in the spectrograms,
fig. \ref{fig:spectrograms}), but the subdominant peaks amplitude in
the full spectrum is lower, decreasing the mass ratio.

\begin{figure*}
  \begin{centering}
    \includegraphics[width=0.95\textwidth]{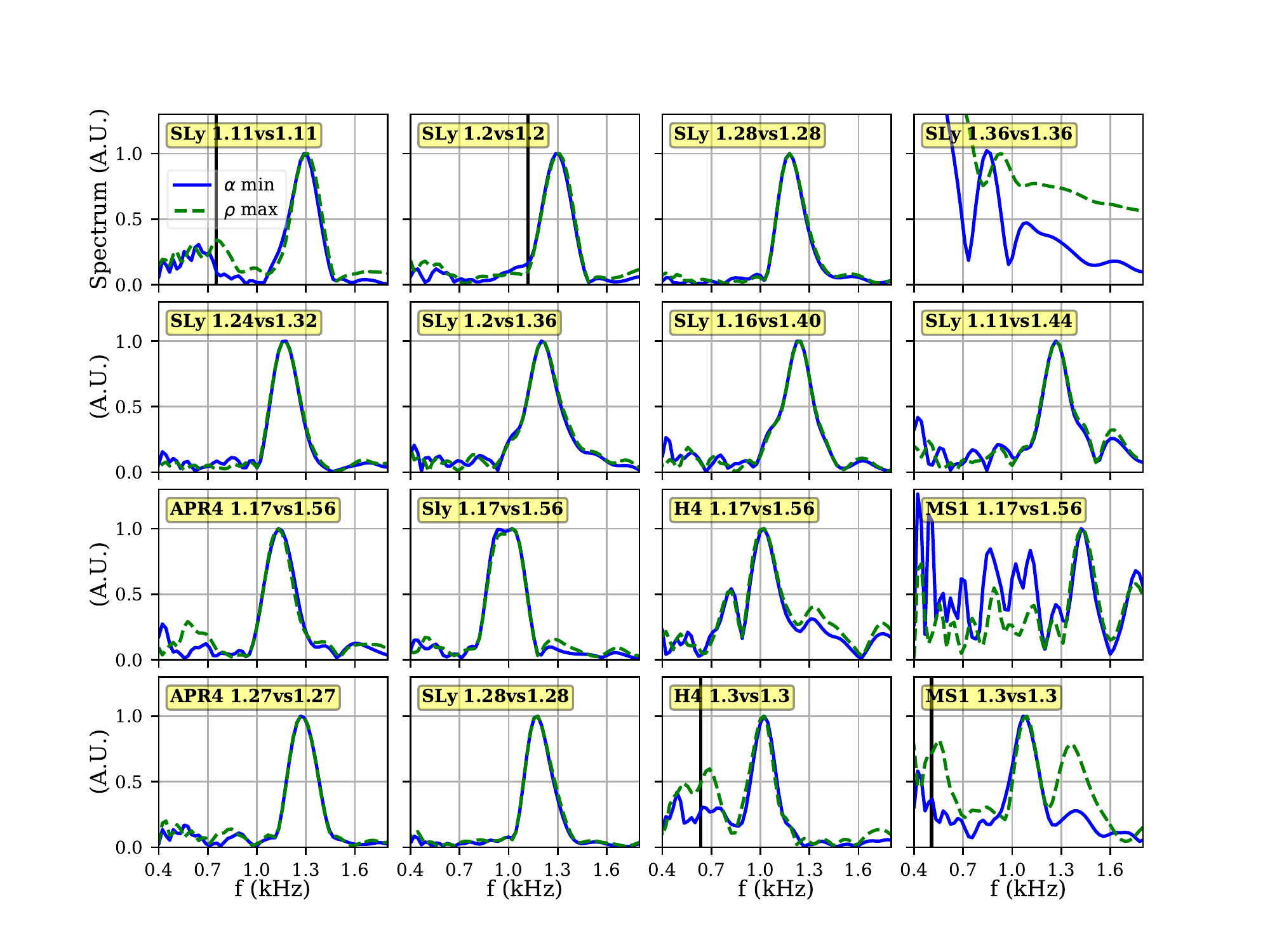}
  \end{centering}
  \vspace{-7mm}
  \caption{Fourier spectra of the maximum density and minimum lapse
    oscillations, computed between the merger and $10$ ms after
    it. The black vertical lines mark, in models for which the
    subdominant GW spectral peaks are not well explained by $m=2$ and
    $m=0$ mode combination, the frequency $f_{2i}-f_1$, at which a
    modulation in the quasi-radial oscillations. In these cases $f_1$
    corresponds to $f_\mathrm{spiral}$, as hypothesized in
    \cite{bauswein:2015unified}}
  \label{fig:rho_spectrum}
\end{figure*}

In ref. \cite{bauswein:2015unified} it was claimed that the GW
emission mechanism at $f_{spiral}$ would leave an observable imprint
also in the maximum density evolution, as a modulation with frequency
$f_{2i} -f_{spiral}$, due to the relative instantaneous orientation of
the external spiral structure respect to the internal double core
structure. To investigate it in our data, we
computed the Fourier spectrogram of the maximum density and minimum
lapse oscillations, in the interval from the merger to $10$ ms after
it. The results are shown in figure \ref{fig:rho_spectrum}, in
arbitrary units, normalized to the spectrum maximum, in order to be
able to compare the spectral features of the two observables. 
In all models, both the maximum density and the minimum lapse spectra show a
dominant peak, corresponding to their oscillation frequency $f_0$. The
peaks in both spectra are remarkably always at the same frequency,
except for the collapsing model with SLy EOS and $M = 1.36 M_{\odot}$
for each star, confirming their ability to measure the quasi-radial
oscillations frequency. While the $\alpha$ spectrum does not show any
other feature, besides the main peak, in some models, in particular,
the less compact ones, the $\rho$ spectrum shows also some subdominant
peaks. This difference between the two observables can easily be
explained by the fact that, while soon after the merger the star lapse
profile has only one single maximum located at the star center, the
density keeps a double core structure, with two rotating local maxima,
for several milliseconds. For the models where $f_{spiral}$ could be a
promising explanation for the low frequency subdominant peak $f_1$ in
the GW spectrum, we drew a vertical, black line at frequency $f_1 -
f_{2i}$ in figure \ref{fig:rho_spectrum}. In the SLy 1.11vs1.11,
H4 1.3vs1.2,MS1 1.3vs1.3 models, it it falls close to a local maximum
in the density spectrum, compatible with it within the Fourier
transform frequency error of $60$ Hz, confirming the presence of a
modulation at that frequency, as predicted in
\cite{bauswein:2015unified}. This finding seems to corroborate the
$f_{spiral}$ hypothesis for those models. In the SLy 1.2vs1.2 and
SLy 1.28vs1.28 models, which showed subdominant emissions in their
spectrograms close both to the prediction for $f_{spiral}$ and to
$f_{2i}-f_0$, instead, only the main $f_0$ peak is present in the
maximum density spectrum.

\subsection{Analyzing the postmerger GW spectrum with the Prony method}
\label{sec:prony}
Fourier spectrograms are a very informative technique, but they have
also some known drawbacks:
\begin{itemize}
\item They consider only the modulus of the GW strain,
  not the full complex number;
\item They do not allow extraction of information about the excited mode
  damping times, nor do they indicate whether the modes are growing or vanishing in
  a particular time interval;
  \item Their accuracy in time and frequency is limited by the
    bandwidth theorem.
\end{itemize}
All these shortcomings can be overcome by a complementary analysis
which fits the time domain signal to
a sum of complex exponentials. These can, at each sampled point $n \in
\sq{1,N}$, be expressed as ($T$ is the sampling time) :
\begin{equation}
h[n] = \sum_{k=1}^{M}{A_k e^{(-\frac{1}{\tau_k} + 2\pi i f_k ) T\; n  + i \phi_k}} =
\sum_{k=1}^M{c_k z_k^{n}}, \label{eq:prony}
\end{equation}
where $M$ is the number of signal components, which \textit{a priori}
could be unknown. Retrieving the same
number of excited modes with the same frequencies using this technique
and using a Fourier spectrum would be a confirmation
that the postmerger GW signal is indeed the superposition
of different exponentially-decaying excited modes.

Fitting a sum of complex exponentials with the standard least-square
technique is known to be problematic, for the large number of free
parameters and the sensitivity to the needed initial guess. Moreover,
the number of excited modes must be chosen {\it a priori}
with standard fitting techniques, as adopted for example in
\cite{hotokezaka:2013remnant,Bauswein:2015vxa}.

A different class of techniques, descending from Prony's method, is known in
the signal processing literature to be a solution to these problems of
least-square fitting a sum of complex
exponentials~\cite{Hua1990,Sarkar1995,Potts2010}.
These methods have already been adopted in
numerical relativity to extract quasi-normal modes from the
ring-down signal in binary black holes simulations~\cite{Berti:2007dg,Berti:2007inspiral},
but have never before been adopted to
study the BNS postmerger GW signal. In particular, we implemented the
ESPRIT Prony variant~\cite{Potts2013,Plonka2014}, which is able to
reconstruct the signal features even in the presence of noise. Like
many other modern Prony implementations, it is based on
fitting a sum of a number of complex exponentials $L$ much larger than
the $M$ ones present inside the signal, and then discriminating
between the physical modes and the ones due to noise. Following a
common choice, we used $L = N/3$ in this work.

Prony's original method was developed to fit a noiseless signal of
$N=2M$ samples. Its starting point was rewriting eq. (\ref{eq:prony}) 
as a Vandermonde linear system:
\begin{eqnarray}
\left(
  \begin{array}{cccc}
 z_1^0 & z_2^0 & \cdots & z_{M}^0\\
 z_1^1 & z_2^1 & \cdots & z_{M}^1\\
  \vdots & \vdots & \ddots & \vdots\\
  z_1^{M\text{-}1} &z_2^{M\text{-}1}&\cdots&z_{M}^{M\text{-}1}
  \end{array}
\right) \left(
\begin{array}{c}
c_1 \\
c_2\\
\vdots \\
c_{M}\\
\end{array}
\right)
= \left(
\begin{array}{c}
h[0] \\
h[1]\\
\vdots \\
h[\mbox{M\text{-}1}]\\
\end{array}
\right)
{}\,.\label{eq:Vandermonde}
\end{eqnarray}
The goal of Prony's method is to find in an independent way a
solution for the complex exponentials $z_k$, which give the
frequencies and the damping times of the signal components. Once they
are known, one can solve system~\ref{eq:Vandermonde} with standard
techniques, to get also the amplitudes and the phases encoded in the
coefficients $c_k$. The starting point is to construct an M-grade
polynomial, whose zeros are the first M $z_k$:
\begin{equation}
p(z) = \prod_{k=1}^{M}{(z - z_k)} = \sum_{k=0}^{M-1}{a_k z^k} + z^M,
z\in \mathcal{C}, \label{eq:pronypoly}
\end{equation}
where the coefficient $a_M$ has been arbitrarily set to one. Starting
from these \textit{Prony polynomials}, one can find the following
relation, for each $m \in \mathcal{N}^*$:
\begin{align}
&\sum_{k=0}^{M}{a_k h[k+m]}\ =\ \sum_{k=0}^M{a_k \rnd{\sum_{j=1}^M{c_j
      z_j^{k+m}}}} = \notag \\ &= \sum_{j=1}^M{c_j z_j^m
  \rnd{\sum_{k=0}^M{a_k z_j^k}}}\ =\ \sum_{j=1}^M{c_j z_j^m p(z_j)}
\ =\ 0.
\end{align} 
Using the sampled values of the signal $h[k],\ k\in [0,2M-1]$, this
can be translated in a forward linear prediction system:
\begin{equation}
\sum_{k=0}^{M-1}{a_k h[k+m]}\ =\ -h[M+m], m \in [0,M-1],
\end{equation}
which, in matrix form, becomes:
\begin{widetext}
\begin{eqnarray}
\left(
  \begin{array}{ccccc}
 h[0] & h[1] & \cdots & h[M-1]\\
 h[1] & h[2] & \cdots & h[M]\\
  \vdots & \vdots & \ddots & \vdots                   \\
 h[M-1] &h[M]&\cdots&h[2M-2]
  \end{array}
\right) \left(
\begin{array}{c}
a[0] \\
a[1]\\
\vdots \\
a[M-1]\\
\end{array}
\right)
=- \left(
\begin{array}{c}
h[M] \\
h[M+1]\\
\vdots \\
h[2M-1]\\
\end{array}
\right)
{}\,.\label{eq:hankel}
\end{eqnarray}
\end{widetext}
In order to fit a sample with noise and $N > 2M$
points, where $M$ is not known a priori, in the ESPRIT Prony technique, the starting point is
building the rectangular Hankel matrix
\begin{equation}
H(0) = 
\left(
\begin{array}{cccc}
h[0] & h[1] & \cdots & h[L] \\
h[1] & h[2] & \cdots & h[L+1]\\
\vdots & \vdots & & \vdots\\
h[N\text{-}L\text{-}1] & h[N\text{-}L] & \cdots & h[N\text{-}1]
\end{array}
\right)
\end{equation}
and the closely related matrix $H_{N-L,L}(1)$, which is obtained from
$H_{N-L,L}(0)$ by removing the first column and adding a $N-L$ vector of
zeros as the last column. Following from eq.~(\ref{eq:hankel}), an
\textit{extended companion matrix} $C_{L+1}$ can be constructed, which
allows one to transform $H_{N-L,L}(0)$ in $H_{N-L,L+1}(1)$:
\begin{align}
&H_{N-L,L+1}(0)C_{L+1} = H_{N-L,L+1}(1) \label{eq:companion}\\
&C_{L+1} = 
\left(
\begin{array}{cc}
C_M(a) & \mathbf{0}_{M,L+1-M}\\
\mathbf{0}_{L+1-M,M} & V_{L+1-M}
\end{array}
\right),
\end{align}
where $C(M)$ is the companion matrix in the original Prony method:
\begin{equation}
C_M(a)\ =\ \left(
\begin{array}{ccccc}
0 & 0 & \cdots & 0 & -a_0 \\
1 & 0 & \cdots & 0 & -a_1 \\
0 & 1 & \cdots & 0 & -a_2 \\
\vdots & \vdots & & \vdots & \vdots\\
0 & 0 & \cdots & 1 & -a_{M-1}
\end{array}
\right)
\end{equation}
and the bottom-right block is given by:
\begin{equation}
V_{L+1-M}\ =\ \left(
\begin{array}{cc}
\mathbf{0}_{1,L-M} & 0\\
I_{L-M} & \mathbf{0}_{L-M,1}
\end{array}
\right).
\end{equation}
The key of this method is the fact that the companion matrix $C_M$ has
the M complex numbers $z_j,\ j\in [1,M]$ as eigenvalues. Note that in
a noiseless sample, its extended version $C_{L+1}$ has the same $M$
eigenvalues, plus $L+1-M$ additional eigenvalues which are
zero. Therefore, the technique focuses on finding the $M$ significant
eigenvalues of $C_{L+1}$ by discriminating them from the eigenvalues due
to noise.

As in many other Prony-like techniques, this is done by performing a
singular value decomposition (SVD) of the Hankel matrix and the
closely related matrix $H(1)$:
\begin{align}
&H_{N-L,L+1} = U_{N-L} S_{N-L,L+1} W_{L+1} \\
&H_{N-L,L+1}(1) = U_{N-L} S_{N-L,L+1} W_{L+1}(1), 
\end{align}
where $U$ and $W$ are unitary matrices and $S$ is a rectangular
diagonal matrix, whose nonzero values $\sigma_i, i \in [1,L+1]$ are
called the \textit{singular values} of the Hankel matrix, arranged in
a non increasing order. $W(1)$ is, by construction of the Hankel
matrices, built from $W$ by removing the first column and adding a last
column filled with zeros. For noiseless data, only $M$ singular values
are nonzero. For data with noise instead, it is possible to define a
threshold $\epsilon$ depending on the desired accuracy (which depends
also on the input data accuracy), in order to find \textit{a posteriori}
the number $M$ of complex exponential components
present in the signal, requiring
\begin{equation}
\frac{\sigma_M}{\sigma_1} \geq \epsilon. \label{eq:Mcondition}
\end{equation}
In the present case, we have chosen $\epsilon = 10^{-2}$.

After the $\sigma_i$ rearrangement, and after determining the value of $M$,
it is possible to remove all $L+1-M$ singular values
linked with noise from $S$ by setting them to zero, and building the rectangular
diagonal submatrix $S_{N-L,M}$.
The submatrix $W_{M,L+1}$ is defined accordingly.
Those submatrices only take into account the signal-related singular values,
are then used to
reconstruct $H_{N-L,L+1}$ and $H_{N-L,L+1}(1)$, choosing also in
this case only the largest $M$ singular values. This allows one to rewrite
eq.~(\ref{eq:companion}) as:
\begin{equation}
S_{N-L,M} W_{M,L+1} C_{L+1}\ =\ S_{N-L,M} W_{M,L+1}(1).
\end{equation}
Multiplying the conjugate transposed equation with
$\rnd{S^*_{N-L,M}}^{\dagger}$ from the left, and setting
\begin{equation}
W_{M,L}(s)\ =\  W_{M,L+1}(1:M,1+s:L+s), s=0,1,
\end{equation}
in order to remove the zero columns, one finally gets 
\begin{equation}
C_L^* W^*_{M,L}(0) = W^*_{M,L}(1).\label{eq:comp2}.
\end{equation}
Since $C_L$ has rank $M$, and its eigenvalues are the $z_j$ we are
looking for, one can find them solving eq. (\ref{eq:comp2}) in the
least-square sense and computing the eigenvalues of the solution
matrix
\begin{equation}
F_M\ :=\ \rnd{W^*_{M,L}}^{\dagger}(0) W^*_{M,L}(1),
\end{equation}
where $\rnd{W^*_{M,L}}^{\dagger}$ is the Moore-Penrose pseudoinverse
of $W_{M,L}$.  Once one obtains the $M$ complex $z_J$x, as eigenvalues
of $F_M$, it is possible to solve the (now overdetermined) Vandermonde
system~\ref{eq:Vandermonde}, again in the least-squares sense, to get
also the $c_j$, from which the modes amplitudes and phases can be
computed.

\begin{figure*}
  \begin{centering}
    \includegraphics[width=0.95\textwidth]{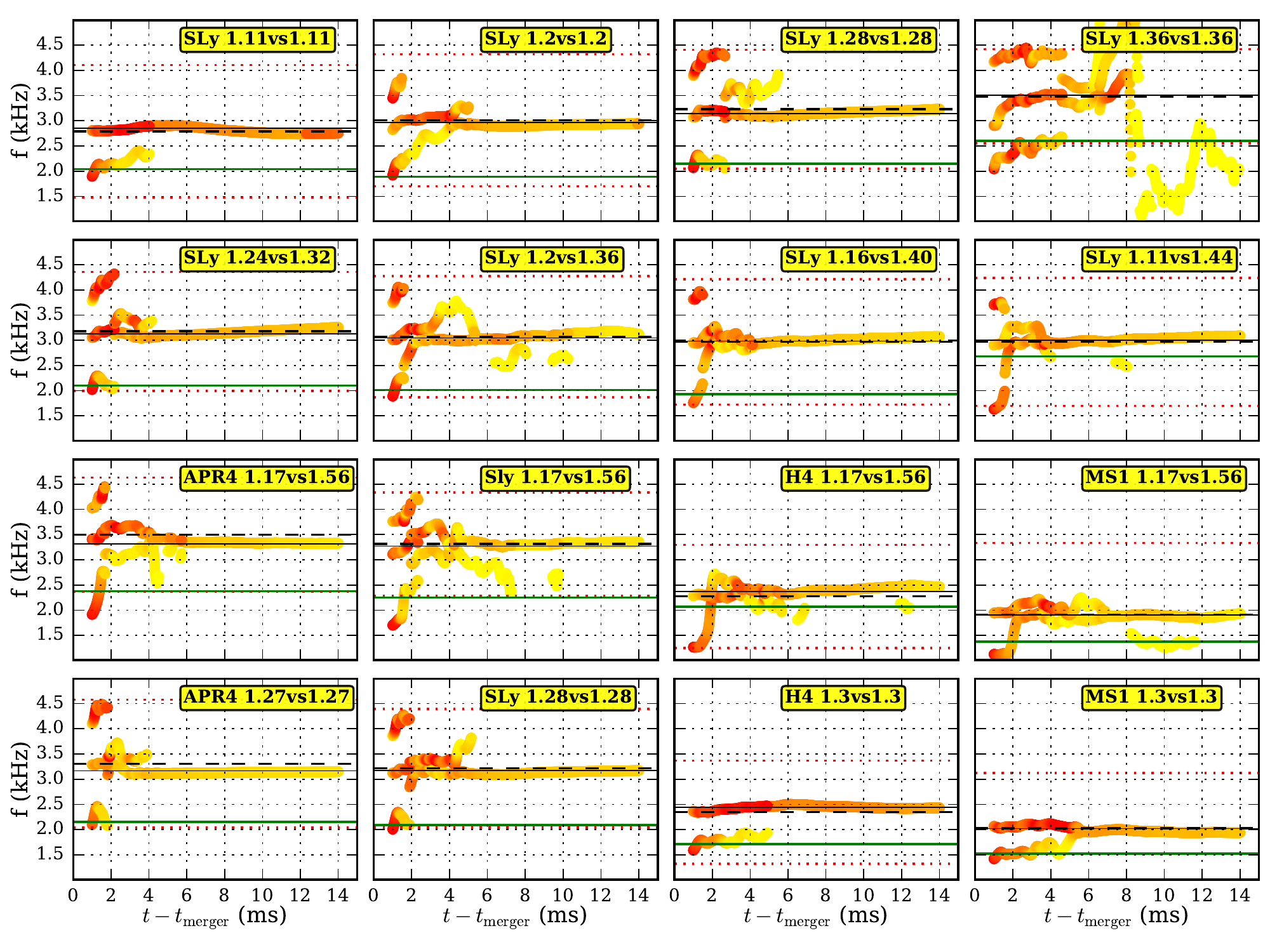}
  \end{centering}
  \vspace{-7mm}
  \caption{Spectrograms constructed applying an ESPRIT Prony algorithm
    in an interval $2$~ms wide around each point. The color code
    refers to the signal component amplitude (darker colors refer to
    higher amplitudes), normalized to the maximum amplitude for a
    component in all the postmerger signals for each model. The
    horizontal lines correspond to the $f_1$ (green), $f_{2i}$ (dashed-black) and 
    $f_{2i}-f_0$ and $f_{2i}+f_0$ (dotted-red) frequencies reported in
    Tables~\ref{tab:f2} and~\ref{tab:f1}.
  \label{fig:Prony}}
\end{figure*}

\begin{table}
  \begin{tabular}{rl@{\hskip 1.5em}ccc}
    \multicolumn{2}{c}{\multirow{2}{*}{Model}} & $f_{1_\mathrm{Prony}}$ & $f_{2i_\mathrm{Prony}}$ & $f_{3_\mathrm{Prony}}$  \\
                                              && [kHz] & [kHz] & [kHz]  \\
    \hline
    SLy &1.11vs1.11  & 2.15 & 2.81 &  -   \\
    SLy &1.20vs1.20  & 1.86 & 2.98 & 3.97 \\
    SLy &1.28vs1.28  & 2.12 & 3.21 & 4.28 \\
    SLy &1.36vs1.36  & 2.34 & 3.40 & 4.33 \\
    \hline          
    SLy &1.24vs1.32  & 2.06 & 3.21 & 4.27 \\
    SLy &1.20vs1.36  & 2.08 & 3.11 & 4.09 \\
    SLy &1.16vs1.40  &  -   & 2.98 & - \\
    SLy &1.11vs1.44  &  -   & 2.93 & - \\
    \hline          
    APR4 &1.17vs1.56 & 2.55 & 3.54 & 4.34\\
    SLy  &1.17vs1.56 & 2.45 & 3.37 & 4.22\\
    H4   &1.17vs1.56 &  -   & 2.24 & - \\
    MS1  &1.17vs1.56 &  -   & 2.00 & -\\
    \hline          
    APR4 &1.27vs1.27 & 2.00 & 3.30 & 4.48\\
    SLy  &1.28vs1.28 & 2.04 & 3.22 & 4.23\\
    H4   &1.30vs1.30 & 1.78 & 2.36 & -\\
    MS1  &1.30vs1.30 & 1.52 & 2.05 & -\\
  \end{tabular}
  \caption{Frequencies of the dominant and subdominant components
    fitted by the ESPRIT Prony algorithm in an interval between $1$~ms
    and $3$~ms after the merger.}
  \label{tab:Prony}
\end{table}

\subsection{Applying Prony's method to our models}
The Prony method is designed to fit signal components with fixed
frequencies. It is therefore important
to perform the Prony analysis only in short
time intervals, as the emission frequencies change with
time, as seen before in the Fourier spectrograms
(figure \ref{fig:spectrograms}).
We applied the ESPRIT Prony algorithm described above to the
postmerger GW strain, in an interval from $1$~ms to $3$~ms after the
merger (when the subdominant modes are active), in order to compute
the component frequencies and compare them to those computed from
the Fourier spectra (reported in tables~\ref{tab:f2} and
\ref{tab:f1}). The results of this analysis are reported in table~\ref{tab:Prony}.
Another useful information that can be extracted using Prony's method
is the determination of the dumping (growing) time associated with a
specific mode of a given fixed frequency. The correct determination of
the dumping time would had required a larger time interval (here we
use a 2 ms window) over which the frequency of the mode is
constant. Unfortunately, the frequency of active emission modes is
changing with time and for this reason the associated damping 
times $\tau$ are much more sensitive to size of the time window and are not 
reported here. On the contrary, it allows to perform time-frequency 
analysis using a small time window maintaining a high precision in the 
determination of the modes frequencies.

Additionally, we built a \textit{Prony spectrogram} for each model
(shown in figure \ref{fig:Prony} where darker colors refer to higher amplitudes), 
assigning the frequencies of the components
fitted by an ESPRIT Prony algorithm applied in a $2$~ms interval
around each point in time for the simulated postmerger evolution.
The points in the plot are colored with a colormap based on
the amplitude of each mode, normalized to the maximum amplitude for
each model. The dashed lines show the values of $f_2$ (black), $f_1$
and $f_3$ (green) from tables \ref{tab:f2} and \ref{tab:f1}. The
components fitted by the Prony method agree well with the Fourier
spectra peaks in most of the simulated models.  The analysis of the 
\textit{Prony spectrogram}, however shows in a precise way that a 
three-peaks structure:
\begin{itemize}
\item it is clearly present only in the two most massive case,
      namely model SLy 1.28vs1,28 and SLy 13.6vs1.36,
\item it is gradually suppressed changing the mass ration $q=M_1/M_2$, and   
\item it does not show up for stiffer equations of state, e.g., in
      models H4~1.3vs1.3 and MS1~1.3vs1.3.
\end{itemize}
This presence of a three-peaks structure
can also be seen in the spectrogram of 
Figure~\ref{fig:spectrograms}, but there  it is by far not as clearly visible as in the 
\textit{Prony spectrogram}.

We also note that, while it might be tempting to see $f_1$ and $f_3$ as equidistant
to $f_2$ when only considering equal-mass models, the unequal-mass models show that
this is not the case there. It is apparent especially from figure~\ref{fig:Prony} (second-row),
that $f_1$ and $f_2$ are not only non equidistant in those cases, but that their
dependency on the mass ratio seems to be similar: the more unequal a system is, the lower
both frequencies seem to be, while $f_2$ hardly changes in comparison. Also, for cases
that show a pronounced frequency $f_3$, the time evolution of both $f_1$ and
$f_3$ seems to show common features, e.g., common raises and drops, while
maintaining a rough factor of two. While such a connection has not been shown rigorously,
it suggests a close physical connection between the processes generating both frequencies,
possibly even being produced by the same process. Understanding why $f_3$ is missing
for especially the low-mass and soft-EOS models could help one understand the process(es)
for both $f_1$ and $f_3$, but due to the need for more model data this remains part
of future work.

\section{Conclusions}
\label{sec:conclusions}

In this paper we studied the GW emission from BNS
merger remnants, focusing in particular on their spectral
features. We analyzed the output of several numerical relativity
simulations of models already considered in our previous works
\cite{DePietri:2015lya,Maione:2016zqz,Feo:2016cbs}, which cover a
relevant portion of the BNS mergers parameter space, varying the
total mass, the mass ratio, and the high-density EOS.

We compared the peak frequency of the Fourier spectra with two
different empirical relations, linking it to static stars
characteristics, and, therefore to the
EOS~\cite{Bauswein:2015vxa,Lehner:2016lxy}, using our data set as an
independent test. The relation from~\cite{Bauswein:2015vxa} showed a
better agreement, with a difference between the predicted and measured
$f_2$ of less than twice the Fourier transform sensitivity in most
equal-mass models (except APR4 1.27vs1.27). The error in the estimated
radius for a static neutron star of reference mass $M=1.6 M_{\odot}$
is at most $350$~m for equal mass models, while it is higher for most
of the simulated unequal-mass binaries. For example, in the sequence
of unequal-mass binaries with fixed EOS (SLy) and total baryonic mass
($M_T = 2.8 M_{\odot}$), the radius error increased by about a factor of
three between $q=0$ and $q=0.83$.

Next, we analyzed the subdominant peaks in the GW spectrum, comparing
different interpretations for their origin. While in most models the
$m=0$ and $m=2$ mode combination hypothesis agrees well with our data,
it does not work for some of the less compact stars (with either low
mass or stiff EOS), for which, additionally, only a low-frequency
subdominant peak $f_1$ is present, with no high-frequency counterpart
$f_3$. In the stiff EOS models, the universal
relation of~\cite{Takami:2014tva,Rezzolla:2016nxn}, which works well also in all
the more compact models, gives a good prediction for the subdominant
peak $f_1$, while it does not work for model SLy 1.11vs1.11. The
$f_{spiral}$ hypothesis of~\cite{bauswein:2015unified,Bauswein:2015vxa}, on the other hand, seems
to work well in all the simulations of our sample where the $f_1$ peak
cannot be explained by mode combination. In particular, in three of
our models (SLy 1.11vs1.11, H4 1.3vs1.3, and MS1 1.3vs1.3)
a subdominant peak at frequency close to $(f_{2i} - f_1)$
is found by analyzing the
spectrum of the maximum density to study the quasi-radial
oscillation. This has been attributed in the cited works to the different
orientation of the outer spiral-arms structure with respect to the inner
double-core structure, which rotates with different angular velocity.

We applied, for the first time in the analysis of GW
from BNS postmergers, a modern variant of Prony's method, which is
a technique to fit a signal with a sum of complex exponentials. This
allowed us to confirm that in the initial transient phase the postmerger GW signal
is indeed a combination of different complex
exponential components, whose frequencies were similar to the values
of the spectral peaks. Also, the number of retrieved components, which
is not imposed \textit{a priori} in Prony's analysis,
agreed well with the subdominant peaks we were able to distinguish in
the spectrum, confirming the subdominant peak suppression in unequal
mass models and the absence of a high-frequency peak $f_3$ in the less
compact models, where mode combination does not explain the $f_1$
frequency well.

We also analyzed the time evolution of the frequency-domain signal,
utilizing both Fourier spectrograms and Prony spectrograms.
Both share the same dynamical features:
\begin{itemize}
\item A change in the dominant peak frequency between the initial
  transient phase and the following quasi-stationary phase.
  It is apparent that this transient is not a sudden
  jump, but rather a continuous process, in which the dominant
  frequency first increase and then decrease;
\item A slow increase in the dominant frequency in the
  quasi-stationary phase which, in particular in the Fourier
  spectrograms, seems more pronounced in equal mass binaries and
  suppressed in unequal mass ones.
\end{itemize}

Overall, our analysis reveals that the spectral properties of the 
postmerger gravitational-wave signal are in agreement with those proposed by 
other groups. In particular, four main peaks appear to be present in all 
of our simulations: $f_1$/$f_{spiral}$, $f_2$, $f_3$ and $f_{20}$. We also find 
that the mechanical toy model of \cite{Takami:2014tva} provides an effective 
description of the early stages of the postmerger and that the two 
main subdominant peaks $f_1$ and $f_3$ are produced only during a transient 
stage of a few milliseconds after the merger, as predicted by the toy model. 
At the same time, we find that the analysis of \cite{bauswein:2015unified} is 
particularly effective for binaries with very low masses or small mass ratios, 
where the $f_{spiral}$ peak replaces the $f_1$ peak and provides a 
better match to the data. This suggests that at least two different mechanisms 
should be considered for the physical interpretation.

A complete understanding, possibily with the aid of the toy model, of the common dynamics 
of the subdominant peaks $f_1$ and $f_3$, and why $f_3$ is suppressed for low-mass and 
stiff-EOS models, could help understanding the process(es) that originate the presence (or absence) of 
subdominant peaks. In principle, we expect that merging binaries will have masses around 1.33 
$M_\odot$ and mass ratios around 1. However, we here suggest that the identification of 
the low-frequency subdominant mode is made either with expression (\ref{eq:f1T}) -- which 
is relative to $f_1$ -- or with expression (\ref{eq:fspiralT}) -- which is relative to $f_{spiral}$ and 
may be more accurate for low-mass binaries and small mass ratios.

We note that, as discussed in \cite{Rezzolla:2016nxn},
the main characteristics of the postmerger spectrum are captured by three main peaks $f_1$, $f_2$, $f_3$
that are closely physical related plus an additional peak denoted as $f_{20}$. This general 
picture was used to get information on the EOS by performing the stacking of multiple BNS 
postmerger events \cite{Bose:2017jvk}. The idea of multiple stacking was also considered in 
\cite{Yang:2017xlf} focusing on just the main (later time) $f_2$ mode.

\begin{table*}  
\begin{center}
\begin{tabular}{rlccc@{\hskip 1.5em}cc@{\hskip 1.5em}cc@{\hskip 1.5em}cc@{\hskip 1.5em}cc@{\hskip 1.5em}c}
\multicolumn{3}{c}{Model} & $M^0_{1}$&$M^0_{2}$&$M_{1}$&$C_1$  &  $M_2$ & $C_2$  &   $M$  &  $C$  & $M_\mathrm{ADM}$ & $J_\mathrm{ADM}$ & $\Omega_0$ (krad/s)\\
\hline
SLy &1.11vs1.11   &  M=1.11     & 1.20 & 1.20 & 1.11 & 0.139 & 1.11 & 0.139 & 1.11 & 0.139 & 2.207 & 5.076 & 1.932   \\
SLy &1.20vs1.20   &  M=1.20     & 1.30 & 1.30 & 1.20 & 0.150 & 1.20 & 0.150 & 1.20 & 0.150 & 2.373 & 5.730 & 1.989   \\
SLy &1.28vs1.28   &  M=1.28     & 1.40 & 1.40 & 1.28 & 0.161 & 1.28 & 0.161 & 1.28 & 0.161 & 2.536 & 6.405 & 2.040   \\
SLy &1.36vs1.36   &  M=1.36     & 1.50 & 1.50 & 1.36 & 0.171 & 1.36 & 0.171 & 1.36 & 0.171 & 2.697 & 7.108 & 2.089   \\
SLy &1.44vs1.44   &  M=1.44     & 1.60 & 1.60 & 1.44 & 0.182 & 1.44 & 0.182 & 1.44 & 0.182 & 2.854 & 7.832 & 2.134   \\
\hline 
SLy &1.24vs1.32   &  q=0.94     & 1.35 & 1.35 & 1.24 & 0.155 & 1.32 & 0.166 & 1.28 & 0.161 & 2.536 & 6.397 & 2.040   \\
SLy &1.20vs1.36   &  q=0.88     & 1.30 & 1.50 & 1.20 & 0.150 & 1.36 & 0.171 & 1.28 & 0.161 & 2.535 & 6.376 & 2.040   \\
SLy &1.16vs1.40   &  q=0.83     & 1.25 & 1.55 & 1.16 & 0.145 & 1.40 & 0.177 & 1.28 & 0.161 & 2.533 & 6.337 & 2.040   \\
SLy &1.11vs1.44   &  q=0.77     & 1.20 & 1.60 & 1.11 & 0.139 & 1.44 & 0.182 & 1.27 & 0.160 & 2.531 & 6.281 & 2.039   \\
\hline 
APR4 &1.17vs1.56  &    APR4     & 1.27 & 1.75 & 1.17 & 0.153 & 1.56 & 0.204 & 1.37 & 0.179 & 2.708 & 7.238 & 1.816   \\
SLy  &1.17vs1.56  &     SLy     & 1.27 & 1.74 & 1.17 & 0.147 & 1.56 & 0.199 & 1.37 & 0.173 & 2.708 & 7.238 & 1.816   \\
H4   &1.17vs1.56  &      H4     & 1.25 & 1.71 & 1.17 & 0.123 & 1.56 & 0.167 & 1.37 & 0.145 & 2.708 & 7.238 & 1.816   \\
MS1  &1.17vs1.56  &     MS1     & 1.25 & 1.70 & 1.17 & 0.117 & 1.56 & 0.154 & 1.37 & 0.135 & 2.708 & 7.238 & 1.816   \\
\hline 
APR4 &1.27vs1.27  & APR4 (1.27) & 1.40 & 1.40 & 1.28 & 0.166 & 1.28 & 0.166 & 1.28 & 0.166 & 2.528 & 6.577 & 1.767   \\
SLy  &1.28vs1.28  & SLy (1.28)  & 1.40 & 1.40 & 1.28 & 0.161 & 1.28 & 0.161 & 1.28 & 0.161 & 2.538 & 6.623 & 1.770   \\
H4   &1.30vs1.30  & H4 (1.3)    & 1.40 & 1.40 & 1.30 & 0.137 & 1.30 & 0.137 & 1.30 & 0.137 & 2.576 & 6.802 & 1.783   \\
MS1  &1.30vs1.30  & MS1 (1.3)   & 1.40 & 1.40 & 1.30 & 0.129 & 1.30 & 0.129 & 1.30 & 0.129 & 2.585 & 6.850 & 1.787   \\
\end{tabular}
\end{center}
\caption{Properties of the analyzed models. Here $M^0_1$ and $M_2^0$ are the total baryonic mass 
of the two stars, $M_1$ and $M_2$ are the mass at infinite separation 
of the two stars, $C_1$ and $C_2$ their compactness while $M$ and $C$ are their average mass.
$M_{ADM}$, $J_{ADM}$ are the total mass and angular momentum of the initial data and $\Omega_0$ 
is the initial angular frequency of the binary system: all the quantities are reported in the 
unit system  where $G=c=M_\odot=1$ except $\Omega_0$ that is reported in krad/s.  
\label{tab:SUM}}
\end{table*}

\begin{table*}  
\begin{center}
\begin{tabular}{rl@{\hskip 1.5em}cc@{\hskip 1.5em}ccc@{\hskip 1.5em}c@{\hskip 1.5em}ccc@{\hskip 1.5em}cc@{\hskip 1.5em}cc}
\multicolumn{2}{c}{Model} & $f_2$    & $f_0$  &  $f_1$  &  $f_{2i}$  &  $f_3$   
                          & $f_2^B$
                          & $f_\mathrm{peak}$ & $f_\mathrm{spiral}$ &  $f_{20}$  
                          & $f^T_1$   & $f^T_\mathrm{spiral}$ 
                          & $f^R_1$   & $f^R_2$ \\
\hline
SLy &1.11vs1.11    & 2.852  & 1.311 & 2.04  & 2.791 &       & 2.784 & 2.257 & 1.657 & 1.241 &  1.641 & 2.041 &  1.657 & 2.360  \\
SLy &1.20vs1.20    & 2.964  & 1.311 & 1.89  & 3.009 & 4.17  & 3.009 & 2.598 & 1.905 & 1.531 &  1.798 & 2.176 &  1.804 & 2.664  \\
SLy &1.28vs1.28    & 3.139  & 1.178 & 2.15  & 3.227 & 4.30  & 3.212 & 2.957 & 2.215 & 1.887 &  1.966 & 2.333 &  1.958 & 2.921  \\
SLy &1.36vs1.36    & 3.506  & 0.933 & 2.60  & 3.483 & 4.30  & 3.410 & 3.358 & 2.603 & 2.329 &  2.211 & 2.521 &  2.210 & 3.156  \\
\hline
SLy &1.24vs1.32    & 3.131  & 1.178 & 2.10  & 3.175 & 4.26  & 3.210 & 2.953 & 2.212 & 1.883 &  1.965 & 2.331 &  1.957 & 2.919  \\
SLy &1.20vs1.36    & 3.047  & 1.200 & 2.01  & 3.072 & 4.19  & 3.210 & 2.956 & 2.214 & 1.886 &  1.966 & 2.333 &  1.958 & 2.920  \\
SLy &1.16vs1.40    & 2.984  & 1.245 & 1.93  & 2.967 & 4.11  & 3.210 & 2.959 & 2.217 & 1.890 &  1.968 & 2.336 &  1.960 & 2.922  \\
SLy &1.11vs1.44    & 2.998  & 1.267 &       & 2.967 &       & 3.197 & 2.943 & 2.202 & 1.872 &  1.959 & 2.332 &  1.951 & 2.911  \\
\hline
APR4 &1.17vs1.56   & 3.317  & 1.133 & 2.36  & 3.494 & 4.62  & 3.574 & 3.657 & 2.914 & 2.681 &  2.449 & 2.756 &  2.478 & 3.305  \\
SLy  &1.17vs1.56   & 3.270  & 1.022 & 2.25  & 3.312 & 4.23  & 3.427 & 3.418 & 2.665 & 2.399 &  2.256 & 2.559 &  2.258 & 3.188  \\
H4   &1.17vs1.56   & 2.370  & 1.022 & 2.06  & 2.274 &       & 2.503 & 2.443 & 1.785 & 1.391 &  1.730 & 1.758 &  1.743 & 2.535  \\
MS1  &1.17vs1.56   & 1.902  & 1.422 & 1.39  & 1.913 &       & 2.179 & 2.171 & 1.606 & 1.179 &  1.594 & 1.539 &  1.606 & 2.271  \\
\hline
APR4 &1.27vs1.27   & 3.166  & 1.267 & 2.16  & 3.307 & 4.48  & 3.336 & 3.155 & 2.403 & 2.101 &  2.078 & 2.513 &  2.069 & 3.043  \\
SLy  &1.28vs1.28   & 3.172  & 1.178 & 2.12  & 3.219 & 4.35  & 3.212 & 2.957 & 2.215 & 1.887 &  1.966 & 2.333 &  1.958 & 2.921  \\
H4   &1.30vs1.30   & 2.448  & 1.023 & 1.71  & 2.349 &       & 2.382 & 2.218 & 1.633 & 1.212 &  1.620 & 1.674 &  1.635 & 2.320  \\
MS1  &1.30vs1.30   & 2.023  & 1.089 & 1.52  & 2.031 &       & 2.081 & 2.024 & 1.530 & 1.087 &  1.494 & 1.506 &  1.489 & 2.100  \\
\end{tabular}
\end{center}
\caption{For each model are report (in kHz) the computed frequency from the simulations ($f_2$, $f_0$, $f_1$, $f_{2i}$, $f_3$
defined in the main text) and the one derived using the proposed universal relations discussed in the appendix. All the frequencies
are expressed in kHz. 
\label{tab:FREQ}}
\end{table*}

\acknowledgments

This project greatly benefited from the availability of public software that
enabled us to conduct all simulations, namely ``LORENE'' and the ``Einstein
Toolkit''. We do express our gratitude to the many people that contributed to
their realization. We would like to thank Andreas Bauswein, Jos\'e 
Antonio Font, Luciano Rezzolla and Nikolaos Stergioulas, for hints, 
discussions and insight that improved the quality of this work.
This work would not have been possible without the CINECA-INFN agreement that
provides access to resources on GALILEO and MARCONI at CINECA. 
Other computational resources were provided 
by he Louisiana Optical Network Initiative (QB2, allocations loni\_hyrel, loni\_numrel,
and loni\_cactus), by the LSU HPC facilities (SuperMic, allocation
hpc\_hyrel) and by PRACE Grant No.\ Pra14\_3593. FL is directly supported by, and this project 
heavily used infrastructure developed using support from the National 
Science Foundation in the USA  (Grants No. 1550551, No. 1550461, No. 1550436, 
No. 1550514).
Partial support from INFN ``Iniziativa Specifica TEONGRAV'' and by the ``NewCompStar'', 
COST Action MP1304, are kindly acknowledged.

\appendix*

\section{Properties of the analyzed Binary Neutron Stars Systems}
\label{sec:appendix}

We report in table~\ref{tab:SUM} the main physical properties of all analyzed models. 
For each model, are reported the total 
baryonic mass of the two stars ($M_1^0$ and $M_2^0$), their gravitational mass and compactness at 
infinite separation and the average gravitational mass $M$ and compactness $C$. In the last 
three columns we report the total mass ($M_{ADM}$), angular momentum ($J_{ADM})$,
and rotational frequency of the initial data, respectively.
In table~\ref{tab:FREQ} we report and summarize the main computed frequency that we derived from the simulations and
the results of the applications of some of the proposed universal relations for the main
frequency of the postmerger spectrum. 

The details of the proposed universal relations reported in table~\ref{tab:FREQ} are the following. 
For equal mass binaries, Bauswein~et.~al.~\cite{Bauswein:2015vxa} proposed the following
universal relation for the main frequency of the post merger with respect to the total mass $M_{tot}=2 \cdot M$ 
(in units of solar mass) of the binary system and the properties of the EOS parametrized as the radius of
the corresponding TOV star of mass $M=1.6 M_\odot$ denoted as $R_{1.6}$ (expressed in km). 
The proposed quasi universal formula is: 
\begin{equation}
f_2^B  = M_{tot} \left( 0.0157 R_{1.6}^2 - 0.5495 R_{1.6} + 5.5030 \right)
\label{formula:f2B}
\end{equation}
The predictions of the unified model by Bauswein and Stergioulas~\cite{bauswein:2015unified} are:
\begin{eqnarray}
f^{U}_{peak}   & = 2.33 - 28.1 \cdot  C + 199 \cdot  C^2\\
f^{U}_{spiral} & = 6.16 - 82.1 \cdot  C + 358 \cdot  C^2 \label{eq:fspiralB}\\
f^{U}_{20}     & = 5.95 - 88.3 \cdot  C + 392 \cdot  C^2
\end{eqnarray}
Analogously we can also check the formula assumed by Rezzolla and Takami \cite{Rezzolla:2016nxn}
\begin{eqnarray}
 f_1^T        &=& -22.07 + 466.62\cdot C  - 3131.63 \cdot C^2  \nonumber \\
              & & + 7210.01\cdot C^3   \; . \label{eq:f1T}
\end{eqnarray}
In \cite{Rezzolla:2016nxn} was also suggested that the third peak $f_3$ is related 
to $f_1^T$ from the prescription $f_3^T=2 f_{2i}-f_1$  and that Eq.~(\ref{eq:fspiralB}) 
should be improved through  a quadratic two-dimensional fit in terms of the compactness and average 
gravitational mass of the binary:
\begin{eqnarray}
 f_{spiral}^T &=&   3.28 -   8.68\cdot C +  174\cdot C^2 \label{eq:fspiralT}\\
              & &  - 2.34 \cdot M + 0.99\cdot M^2 - 13.0\cdot C \cdot M 
   \; .\nonumber 
\end{eqnarray}
More recently, in order to determine the neutron star radius from a population of BNS
mergers~\cite{Bose:2017jvk}, the following fit was proposed for the determination
of the $f_1$ and $f_2$ frequencies as a function of the compactness (since
here we are also considering unequal mass binaries, we use the average compactness):
\begin{eqnarray}
  f_1^R &=& -35.17 + 727.99 \cdot C - 4858.54 \cdot C^2 \nonumber \\
        & &  + 10989.88 \cdot C^3 \\
  f_2^R &=& -3.12  + 51.90 \cdot C - 89.07 \cdot C^2 
\end{eqnarray}
In general, we find that all these formulas show some agreement with the observed 
frequencies within a discrepancy at most of 0.2--0.3 kHz that is not much 
greater than the half-amplitudes of the peaks.

\end{document}